\documentclass[12pt]{article}
\usepackage[reqno]{amsmath}
\usepackage{amssymb, theorem, enumerate}
\usepackage{graphicx}
\usepackage{tikz}
\usetikzlibrary{decorations.markings}
\usepackage{array}
\usepackage{float}
\usepackage{wrapfig}
\usepackage{fullpage}
\usepackage{nicefrac}

\expandafter\ifx\csname\endcsname\relax\else
\pdfpagebox
\fi

\theoremstyle{change}
{\theorembodyfont{\slshape}
\newtheorem{theorem}{Theorem.}[section]
\newtheorem{proposition}{Proposition.}[section]
\newtheorem{lemma}[theorem]{Lemma.}
\newtheorem{corollary}[theorem]{Corollary.}}

\theorembodyfont{\rmfamily}

\newcommand\cref[1]{Corollary~\ref{cor:#1}}

\def\proof{\noindent{{\sl Proof. }}}
\def\sqr#1#2{{\vbox{\hrule height.#2pt
    \hbox{\vrule width.#2pt height#1pt \kern#1pt
        \vrule width.#2pt}\hrule height.#2pt}}}
\def\eqed{\sqr53}
\def\qed{%
    \ifmmode\eqno\eqed
    \else\nobreak\ \hfill\eqed\medbreak\fi}

\newcommand\cA{{\mathcal A}}

\newcommand\cH{{\mathcal H}}

\newcommand\cM{{\mathcal M}}

\newcommand\Zf{{\mathbf f}}
\newcommand\Ze{{\mathbf e}}

\newcommand\Zu{{\mathbf u}}
\newcommand\Zv{{\mathbf v}}
\newcommand\Zz{{\mathbf z}}
\newcommand\Zx{{\mathbf x}}
\newcommand\Zw{{\mathbf w}}

\newcommand\ones{{\mathbf 1}}
\newcommand\zeros{{\mathbf 0}}

\newcommand\cx{{\mathbb C}}

\newcommand\re{{\mathbb R}}

\newcommand\Mat[3]{\text{Mat}_{#1 \times #2}(#3)}

\newcommand\comp[1]{{\mkern2mu\overline{\mkern-2mu#1}}}

\newcommand\pmat[1]{\begin{pmatrix} #1 \end{pmatrix}}

\DeclareMathOperator{\tr}{tr}

\title{Entanglement of Free Fermions on Hadamard Graphs}
\author{Nicolas Cramp\'{e}\footnote{Institut Denis-Poisson CNRS/UMR 7013 - Universit\'e de Tours - Universit\'e d'Orl\'eans, 
Parc de Grandmont, 37200 Tours, France.} , 
Krystal Guo\footnote{Centre de recherches math\'ematiques, Universit\'e de Montr\'eal,
P.O. Box 6128, Centre-ville Station,
Montr\'eal (Qu\'ebec), H3C 3J7, Canada.} , 
Luc Vinet\footnotemark[2]
}

\begin{document}

\maketitle

\begin{abstract}
Free Fermions on vertices of distance-regular graphs are considered.  Bipartition are defined by taking as one part all vertices at a given distance from a reference vertex.
The ground state is constructed by filling all states below a certain energy. Borrowing concepts from time and band limiting problems, algebraic Heun operators and Terwilliger algebras, it is shown how to obtain, quite generally, a block tridiagonal matrix that commutes with the entanglement Hamiltonian.
The case of the Hadamard graphs is studied in details within that framework and the existence 
of the commuting matrix is shown to allow for an analytic diagonalization of the restricted two-point correlation matrix and hence for an explicit determination of the entanglement entropy.
\end{abstract}

\section{Introduction}\label{sec:quantum-intro}
As a probe in particular of the correlations in quantum many body systems and field theories, the study of entanglement is of fundamental interest. We shall here be concerned with free Fermions on graphs, that is we shall explore systems of Fermions hopping between the vertices of distance-regular graphs with  dynamics controlled by the adjacency matrix taken as the Hamiltonian. We shall develop general results and focus on their realization in Hadamard graphs. A key finding will be the identification of a block tridiagonal matrix that commutes with the entanglement Hamiltonian. As will be shown, this will allow for an analytic diagonalization of the chopped correlation matrix on Hadamard graphs. The commuting operator will be found by extending the algebraic Heun operator construction to the framework of the Terwilliger algebra arising in the context of association schemes. This will generalize previous entanglement studies of free Fermion chains \cite{Peschel04, EP04, Eisler18, CNV1, CNV2}, that exploited ideas borrowed from time and band limiting problems \cite{Slepian83, Landau, SP61,G, GVZ1}.  

To talk about entanglement, one needs to take the system under consideration in a given state and to examine how a part of it is correlated to the rest. Often this is done in the ground state. In the case of Fermionic systems, this requires a certain filling of the Fermi sea, there is hence a limited set of energies that is thus selected. Determining a part of the systems in turn involves picking an ensemble of vertices.

Because of their simplicity, free chains have been fairly well studied. In these situations, owing to Wick's theorem, the reduced density matrix can be expressed in thermodynamic form in terms of a so-called entanglement Hamiltonian which is determined by the two-point correlation matrix restricted to the chosen part of the system (for reviews see \cite{LR, Peschel12}). Quite remarkably in some instances it has proven possible to find a tridiagonal matrix that commutes with the correlation matrix and thus shares with it joint eigenvectors. Either for numerical computation issues or asymptotic analyses, this fortuitous circumstance has proved important. 

The existence of this commuting operator for free Fermionic chains could be suspected from the parallel with problems of time and band limiting in signal processing where one wishes to optimize the concentration in time of signals comprised of a limited frequency interval. (The restrictions in the ranges of two dual variables is at the root of the analogy.) The seminal work in the signal analysis context was done by Slepian and coworkers \cite{Slepian83, Landau, SP61}, who identified such a useful operator commuting with the limiting operator. It should be said that this feature has emerged in various other contexts among which random matrix theory or integrable models.

The bearing of all this on the examination of entanglement in free Fermionic chains was brought to light by Eisler and Peschel (among others see also \cite{GK}) who found for example in \cite{Eisler18}, the tridiagonal matrix that commutes with the truncated correlation matrix of a uniform finite chain and studied in detail how this commuting matrix compares with the entanglement Hamiltonian and how it may yield the spectrum of the latter.

The reasons behind the rather miraculous existence of the commuting operators have intrigued many and Grünbaum who has devoted much effort to understand this phenomenon has tied it to the field of bispectral problems that he has established \cite{DG}. This later led to the concept of algebraic Heun operator introduced in \cite{GVZ1,GVZ2} (see also \cite{NT, RKP}) which was shown to naturally provide commuting operators. These methods were applied in \cite{CNV1,CNV2} to various free Fermionic finite chains with non-uniform couplings and possessing a bispectral underpinning. More precisely, they were used to analyse chains described by a hopping matrix $h$  tridiagonal in the characteristic (site) basis and admitting another "position" operator $\mathcal{X}$ tridiagonal in the eigenbasis of $h$. (Such a set of operators conforms to the definition a Leonard pair \cite{T03}.) In situations where the limiting in energy and space involved sequences of consecutive values beginning with the initial one, the algebraic Heun operator - a bilinear expression in the two Leonard operators - was readily found in \cite{CNV1,CNV2} to give the commuting matrix.

We here extend this study to systems of free Fermions on graphs and in particular on those of Hadamard. While the entanglement of such Fermionic configurations has been looked at in \cite{JES1, JES2}, our main interest will be to identify the commuting matrices in the case of distance-regular graphs. This will prove possible when the limitings will respect the distance structure of the graph. Because of the presence of degeneracies, it will call for an extension of the algebraic Heun construction to the framework of Terwilliger algebras \cite{Ter92} which have been introduced to study the properties of association schemes. Simply put these algebras are generated by the adjacency matrix which plays in our models the role of the Hamiltonian and the orthogonal projectors $E^*_i$ on subspaces spanned by the characteristic vectors corresponding to the vertices at distance $i$ from the reference one. (For an introduction to Terwilliger algebras see \cite{GO}.) This will lead to commuting operators that are block-tridiagonal (instead of being simply tridiagonal in the case of the Fermionic chains described in the preceding paragraph). This should be put in parallel with the recent investigations of 
Grünbaun and collaborators to identify commuting operators stemming from matrix orthogonal polynomials \cite{GPZ}.

The analysis will be carried out in details for Hadamard graphs and the usefulness of the commuting block-tridiagonal matrix will prove manifest in this case. The Hadamard graph of order $n$ has diameter $4$ and $4n$ vertices \cite{BCN}; their set form a self-dual association scheme. The spectral properties of Hadamard graphs make them a prime candidates, amongst distance regular graphs, for a variety of applications; for example they are studied in the context of state transfer in quantum walks, see \cite{CoutinhoGodsilGuoVanhove2}. Given a reference vertex, the chopped correlation matrix will be expressed in terms of projectors made out of the primitive and dual idempotents of the scheme; it will therefore belong to the associated Terwilliger algebra. One usually want to obtain the eigenvalues of the correlation matrix to determine for instance the entanglement entropy. The essential merit of the commuting operator that shares common eigenvectors with the correlation matrix is that in contrast to the latter, it generally lends itself to a tractable numerical diagonalization. Here, quite remarkably, for Hadamard graphs, it will be seen to allow in fact for an analytic determination of the eigenvectors.

The paper will now unfold as follows. Section 2 will provide the description of free Fermions on graphs. The Hamiltonian will be diagonalized with the help of the primitive idempotents which will also be used to express the two-point correlation matrix in the ground state. Section 3 frames the discussion of entanglement. The system is split in two parts with one comprising the complete neighbourhoods that are consecutive to the reference vertex. That the entanglement Hamiltonian is determined by the restricted correlation matrix is reviewed and this central entity is seen to be given as the product of projectors. Attention will be focused on graphs that belong to association schemes that are P- and Q- polynomials. These will be surveyed in Section 4 where the Terwilliger algebras will be defined. Section 5 contains one of the main results and shows how a block-tridiagonal operator that commutes with the restricted correlation matrix is obtained from the algebraic Heun construction and the Terwilliger algebra. Section 6 shows in detail how the analysis applies in the case of Hadamard graphs. Free Fermions on these graphs will be shown to provide a system where the entanglement Hamiltonian can be diagonalized analytically thanks to the existence of the commuting operator. Exact expressions for the Von Neumann entropies will be obtained as a result. Concluding remarks will be found in the last section.

\section{Free Fermions on graphs}

Let $G$ be a graph, $V=\{v_1,\dots, v_N\}$ the set of its vertices 
and $E$ the set of its edges. 
We define a vector space of dimension $N$ with 
the orthonormal canonical basis $\{ |1\rangle, \dots ,|N\rangle\}$, called position basis, where the vector $|i\rangle$ is associated to the vertex $v_i$.  
We define the adjacency matrix $A$ as a $N\times N$ matrices with the components, for $1\leq i,j \leq N$,
\begin{equation}
   \langle i|A |j\rangle= A_{ij}=\begin{cases}1 & \text{if } (v_i,v_j)\in E\\
    0 & \text{otherwise}
    \end{cases}\,.
\end{equation}
We restrict ourselves to non-oriented edges \textit{i.e.} $(v_i,v_j)\in E \Leftrightarrow (v_j,v_i)\in E$. In this case, the matrix $A$ is symmetric.
We introduce the fermionic operators $c_n$ and $c_{n}^{\dagger}$ ($n=1,2,\dots,N$) satisfying
\begin{equation}
   \{ c_{m} \,, c_{n} \} = 0\,, \quad   
   \{ c_{m}^{\dagger} \,, c_{n}^{\dagger} \} = 0\,, \quad   
   \{ c_{m}^{\dagger} \,, c_{n} \} = \delta_{m,n}, 
   \qquad\text{for } 1\leq m,n \leq N \,.
\end{equation}
We consider the following open quadratic free-Fermion Hamiltonian on the graph $G$
\begin{equation}\label{eq:Hff2}
{\widehat{\mathcal{H}}}=(c^\dagger_1,\dots,c^\dagger_N)\,  A \begin{pmatrix}c_1\\ \vdots \\c_N\end{pmatrix}=
\sum_{(v_i,v_j)\in E} c_i^\dagger c_j\ .
\end{equation}
It describes the hopping of Fermions on the underlying graph $G$. As usual, in free-Fermion models, to diagonalise the Hamiltonian $\widehat{\mathcal{H}}$, one diagonalises the matrix $A$. As that matrix is symmetric, it can be diagonalised by an orthonormal basis $|\theta_k^i\rangle$ for $k=0,1,\dots,\delta$ and $i=1,2\dots f_k$,  
 \begin{equation}
     A |\theta_k^i\rangle =\theta_k |\theta_k^i\rangle \,,
 \end{equation}
 where $\theta_k$ are the pairwise distinct eigenvalues of $A$
 and $f_k$ is the multiplicity of the eigenvalue $\theta_k$.
 The eigenfunctions $\phi_j(\theta_k^i)$ are the components of the vector $|\theta_k^i\rangle$
 in the position basis:
 \begin{equation}
     |\theta_k^i\rangle=\sum_{j=1}^N \phi_j(\theta_k^i) |j\rangle \,.
 \end{equation}
  We order the eigenvalues such that  $\theta_0< \theta_1< \dots < \theta_\delta$.
 We introduce $E_k$, the primitive idempotent associated
 to the eigenspace with eigenvalue $\theta_k$:
 \begin{equation}
     E_k=\sum_{i=1}^{f_k} |\theta_k^i\rangle \langle\theta_k^i|\,.
 \end{equation}
 Let us remark that $f_k=\text{rank}(E_k)$.
 We define also the following projector on the first $K+1$ eigenspaces:
 \begin{equation}\label{eq:pi2}
     \pi_2(K)=\sum_{k=0}^K E_k\,.
 \end{equation}
 These projectors permit to write the adjacency matrix as follows
 \begin{equation}
A = \sum_{k=0}^{\delta} \theta_{k} E_k\,.
\end{equation}

Having diagonalised $A$, the Hamiltonian $\widehat{\cH}$ can be written as follows 
\begin{equation}
    \widehat{\cH}= \sum_{k=0}^\delta \theta_k \sum_{i=1}^{f_k}  (\tilde c_k^i)^\dagger\ \tilde c_k^i\,,
\end{equation}
where the transformed fermionic operators $\tilde c_k^i$ and $(\tilde c_k^i)^\dagger$ are related to the first ones by
\begin{equation}\label{eq:cct}
  \tilde c_k^i= \sum_{j=1}^N \phi_j(\theta_k^i) c_j\quad\text{and}\qquad(\tilde c_k^i)^\dagger= \sum_{j=1}^N \phi_j(\theta_k^i) c_j^\dagger \,.
\end{equation}
Then, the eigenvalues $\mathcal{E}$ of the Hamiltonian $\widehat{\cH}$ are described
by the set of integers $\epsilon_k \in \{0,1,\dots f_k\}$ and are given by
\begin{equation}
   \mathcal{E}= \sum_{k=0}^\delta \epsilon_k \theta_k\,.
\end{equation}

Let $K\in \{0,1,\dots , \delta\}$ such that
\begin{equation}\label{eq:defK}
    \theta_K <0 < \theta_{K+1} \,.
\end{equation}
The number $K$ can be modified by adding a matrix proportional to the identity at the adjacency matrix $A$.
That corresponds to consider an additional constant magnetic field for the Hamiltonian $\widehat{\mathcal{H}}$.
We can also choose this magnetic field such that the inequalities in \eqref{eq:defK} are always strict which simplifies the latter discussion about the ground state and the entropy.
The ground state of the Hamiltonian $\widehat{\cH}$ is 
\begin{equation}\label{eq:GS}
    |\Psi_0\rangle\!\rangle= (\tilde c^1_0)^\dagger\dots (\tilde c^{f_0}_0)^\dagger  \dots (\tilde c^1_K)^\dagger\dots (\tilde c^{f_K}_K)^\dagger|0\rangle\!\rangle
\end{equation}
where the vacuum state $|0\rangle\!\rangle$ is annihilated by the fermionic operators $\tilde c_k^i$:
\begin{equation}\label{eq:vac}
    \tilde c_k^i|0\rangle\!\rangle=0\,,\quad k=0,1,\dots,\delta\,,\quad i=1,2,\dots,f_k\,.
\end{equation}

The correlation matrix $\widehat{C}$ in the ground state $|\Psi_0\rangle\!\rangle$ is the $N \times N$ matrix with components, for $1\leq n,m\leq N$,
\begin{equation}
    \widehat{C}_{mn}=\langle\!\langle \Psi_0 |  c_m^\dagger c_n |\Psi_0\rangle\!\rangle
\end{equation}
By using the second relation in \eqref{eq:cct}, we can rewrite the correlation matrix as follows
\begin{equation}
\widehat{C} =\sum_{k=0}^{\delta}\sum_{i=1}^{f_k}
\sum_{p=0}^{\delta}\sum_{j=1}^{f_p}\langle\!\langle \Psi_0 |
(\tilde c_k^i)^\dagger \tilde c_p^j 
|\Psi_0\rangle\!\rangle
|\theta_k^i\rangle \langle\theta_p^j|\,.
\end{equation}
By rewriting the ground state with its definition \eqref{eq:GS}, by using the anticommutation relations of the fermionic operators and the properties \eqref{eq:vac}, one gets 
\begin{equation}
\widehat{C} =\sum_{k=0}^{K}\sum_{i=1}^{f_k} |\theta_k^i\rangle \langle\theta_k^i|= \sum_{k=0}^{K} E_k\,.
\end{equation}
Therefore the correlation matrix corresponds to the projector $\pi_2$ defined by \eqref{eq:pi2}: $\widehat{C} =\pi_2(K)$.

\section{Entanglement entropy}

In order to examine entanglement, we must first define a bipartition
of our free-Fermionic graph. 
We denote by $d(v,w)$ the distance between the two vertices $v$ and $w$ of the graph $G$. From now on, we fix a vertex $v$ of $G$ and, without loss of generality, we take $v=v_1$.
We choose as
subsystem $1$ the vertex $v_1$ and all the vertices at distance at most $\ell$ from $v_1$: 
\begin{equation}
    G_1=\{w\in V\ |\ d(v_1,w)\leq \ell\}\,.
\end{equation}
Let us denote by $N_1$ the cardinal of the previous set \textit{i.e.} the number of vertices in the subsystem $1$.
We shall
find how it is intertwined with the rest of the graph in
the ground state $|\Psi_{0}\rangle\!\rangle$.  To that end, we need the
reduced density matrix 
\begin{equation}
\rho_{1} = \tr_{2} |\Psi_{0}\rangle\!\rangle 
\langle\!\langle \Psi_{0}| \,,
\end{equation}
where the subsystem $2$ is the complement of the subsystem $1$, from which one can compute for instance the von Neumann entropy 
\begin{equation}
S_{1} = -\tr
(\rho_{1} \log \rho_{1})\ .\end{equation}
The explicit computations of this entanglement entropy amounts to finding the eigenvalues of
$\rho_{1}$.

It has been observed that this reduced density matrix $\rho_1$ is determined by
the spatially ``chopped'' correlation matrix $C$ , 
which is the following $N_1 \times N_1$ submatrix of $\widehat C$: 
\begin{equation}
  C =  |\widehat C_{mn} |_ {m,n \in G_1}\,. 
\end{equation}
The argument which we take from \cite{2003JPhA...36L.205P} (see also
\cite{2009JPhA...42X4003P}) goes as follows.  Because the ground state
of the Hamiltonian $\widehat{\mathcal{H}}$ is a Slater determinant,
all correlations can be expressed in terms of the one-particle
functions, i.e. in terms of the matrix elements of $\widehat C$. 
Restricting to observables associated to part 1, since $\langle A \rangle = \tr (\rho_{1}A)$, 
the factorization property will hold according to Wick's theorem if 
$\rho_{1}$ is of the form 
\begin{equation}
\rho_{1} = \kappa \; \exp (-\mathcal{H}) \,,
\label{entH}
\end{equation}
with the entanglement Hamiltonian $\mathcal{H}$ given by
\begin{equation}
\mathcal{H} = \sum_{m,n =1}^{N_1} h_{mn} \, c_{m}^\dagger 
c_{n} \,.
\label{hopping}
\end{equation}
The hopping matrix $h= |h_{mn}|_{1\leq m,n \leq N_1}$ is defined so that
\begin{equation}
C_{mn} = \tr (\rho_{1} \; c_m^\dagger c_n)\,,  \qquad  m\,, n \in \{
1,2,\dots , N_1\},
\label{obs}
\end{equation}
holds, and one 
finds through diagonalisation that 
\begin{equation}
h = \log [(1 - C) /C] \,.
\end{equation}
We thus see that the $2^{N_1} \times 2^{N_1}$ matrix $\rho_{1}$
is obtained from  the $N_1 \times N_1$ matrix $C$ or equivalently, the entanglement Hamiltonian $\mathcal{H}$.

To describe this chopped correlation matrix $C$, we introduce the projector $E_i^*$ which is the diagonal characteristic matrix of the $i$th neighbourhood of $v_1$; that is $E_i^*$ is the diagonal matrix such that
$\langle j| E_i^* |j\rangle = 1 $
if $d(v_j,v_1) = i$ and 
$\langle j| E_i^* |j\rangle = 0 $ otherwise. Note that
\begin{equation}
E_i^*E_j^*=\delta_{ij} E_i^* \quad \text{and}\qquad 
\sum_{i=0}^d E_i^* = I,
\end{equation}
where $d$, the diameter of the graph $G$, is the maximum distance of any vertex from $v_1$ and $I$ is the identity matrix.
Let $\pi_1(\ell)$ denote the projection onto the first $\ell$ neighbourhoods of $v_1$:
\begin{equation}
\pi_1(\ell) = \sum_{s=0}^\ell E_s^* \,,
\end{equation}
and $n_i$ be the number of vertices at distance $i$ from $v_1$. The operator $\pi_1(\ell)$ projects on 
$N_1=\sum_{i=0}^\ell n_i$ sites.
Finally, the chopped correlation matrix can be written as (see for instance  \cite{Lee:2014nra,2012PhRvB..86x5109H})
\begin{equation}
 C =\pi_1(\ell) \pi_2(K) \pi_1(\ell) \ .
 \label{Cpi}
\end{equation}

To calculate the entanglement entropies one therefore has to compute the
eigenvalues of $C$.  As explained in \cite{Peschel04}, this
is not easy to do numerically because the eigenvalues of that matrix
are exponentially close to $0$ and $1$.  This motivates the search for
a matrix $T$ such that
\begin{equation}
 [T,C] =0 \,,
 \label{want}
\end{equation}
and easier to diagonalise numerically than $C$. We denote the commutator bracket by $[\cdot,\cdot]$; that is, $[A,B] = AB- BA$ for matrices $A,B$.  
We shall hence find a $T$ satisfying  \eqref{want}
by looking for a matrix commuting with both projectors:
\begin{equation}
[T,\pi_1(\ell)] = [T,\pi_2(K)]=0 \,,
\label{pi}
\end{equation}
In \cite{CNV1,CNV2}, a such operator $T$, which is a tridiagonal matrix, for free-Fermion on a line has been constructed when $A$, defining the Hamiltonian $\widehat{\cH}$, belongs to a bispectral problem. In this case, the operator $T$ turns out to be a Heun operator. 
Inspired by these results, we consider a special class of graphs whose associations schemes are $P$- and $Q$-polynomials (which we will provide background on in next section) which replaces the notion of bispectrality. Then, we show that in this case an operator $T$ tridiagonal by block satisfying \eqref{pi} exists.

We may observe that the matrix $D$ defined by 
\begin{equation}\label{eq:defD}
    D= \pi_2(K) \pi_1(\ell) \pi_2(K)\,,
\end{equation} would describe a dual entanglement situation where the vacuum state would be filled with excitations by the set $\{0,...,\ell\}$, and the subsystem would consist of the sites in the neighbourhood $\{0,\dots,K\}$. We recall that for $n\times n$ matrices $M$ and $N$, the characteristic polynomials of $MN$ and $NM$ coincide, see Section 11.6 of \cite{P94}. Recalling that $\pi_i(j)$ is idempotent for $i\in\{1,2\}$, we see that $C$ and $D$ both have the same eigenvalues as the following matrix:
\[
\pi_1(\ell)^2 \pi_2(K) = \pi_1(\ell) \pi_2(K) = \pi_1(\ell) \pi_2(K)^2 .
\]
Since $C$ and $D$ have the same non-zero eigenvalues, the entanglement entropies will be the same in these two instances. Such duality generalizes the one studied in \cite{CFG}.

\section{ $P$- and $Q$-polynomial association schemes} 

\subsection{Association schemes and distance regular graphs}\label{sec:assoc-intro}

A \textsl{distance regular graph} is a graph $G$ of diameter $d$ such that there exists constants $p_{ij}^k$ for $i,j,k \in \{0,\ldots, d\}$ such that, for any pair of vertices $x,y$ at distance $k$, there are $p_{ij}^k$ vertices at distance $i$ from $x$ and distance $j$ from $y$. We follow the standard text \cite{BCN} and the dynamic survey \cite{drgsurvey} and defer to them for further background on distance regular graphs.  For $i = 0,\ldots, d$, the \textsl{distance matrix}, denoted $A_i$, is given by
\[
(A_i)_{x,y} = \begin{cases} 1, & \text{ if } d(x,y) = i;\\
0, & \text{else.} \end{cases}
\]
The distance matrix $A_1$ corresponds to the adjacency matrix $A$ used in the previous sections.
The following is the main theorem about the eigenspaces of distance regular graphs and can be found in \cite{BCN}. In the following theorem, let $N$ the number of vertices of $G$ and  $J$ denotes the all ones matrix, $I$ denotes the identity matrix and $\circ$
denotes the Schur product, the entry-wise matrix multiplication.
\begin{theorem}\label{thm:drg}
The matrices $\{A_0,\ldots,A_d\}$ commute pairwise and share $d+1$ eigenspaces. Let $E_0,\ldots, E_d$ be the primitive idempotent projectors onto the shared eigenspaces of $\{A_i\}_{i=0}^d$. There exists constants $q_{ij}^k$ for $i,j,k \in \{0,\ldots, d\}$ such that following hold:
\begin{enumerate}[(i)]
    \item $A_0 = I$  and  $E_0 = \frac{1}{N}J$;   
\item  $\displaystyle \sum_{i=0}^d A_i = J$  and  $\displaystyle \sum_{i=0}^d E_i = I$;  
\item $A_i \circ A_j = \delta_{ij}A_i$  and  $E_i E_j = \delta_{ij}E_i$; and
\item $A_i A_j = \displaystyle\sum_{k=0}^d p_{ij}^k A_k$  and $E_i \circ E_j =\displaystyle \sum_{k=0}^d q_{ij}^k E_k$. \qed
\end{enumerate} 
\end{theorem}

More generally, if $\{A_0, \ldots, A_d\}$ are symmetric matrices such that Theorem \ref{thm:drg} holds, then we say that they are the \textsl{associate matrices} of an \textsl{(commutative) association scheme}. For brevity, we sometimes refer to an commutative association scheme as a \textsl{scheme}.
The constants $p_{ij}^k$ are known as the \textsl{intersection numbers} of the scheme and the constants $q_{ij}^k$ are known as the \textsl{Krein parameters} of the scheme.

The distance matrices of a distance regular graph form the basis of a $(d+1)$-dimensional, semi-simple, commutative subalgebra $\cM$ of the algebra of $N \times N$ matrices over $\cx$, also call the \textsl{Bose-Mesner algebra} of the graph.

Since $\{E_i\}_{i=0}^d$ and $\{A_i\}_{i=0}^d$ are both bases of $\cM$, there exists change of bases matrices between them. The \textsl{eigenmatrices} of the association scheme are  $d+1 \times d+1$ matrices $P$ and $Q$ such that
\begin{equation}\label{eq:changebasisAE}
A_j = \sum_{i = 0}^d P_{ij}E_i\quad \text{and}\qquad E_j = \frac{1}{N} \sum_{i=0}^d Q_{ij} A_i.
\end{equation}
Note that this implies that $\{P_{ij}\}_{i=0}^d$ are the eigenvalues of $A_j$.

Since $\cM$ is closed under multiplication and addition, there is a choice of the ordering of $A_1,\ldots, A_d$ such that $A_i$ is a polynomial in $A_1$ for all $i$. We say the scheme is \textsl{$P$-polynomial} if $A_i$ is a polynomial in $A_1$ of degree $i$, for each $i = 0,\ldots, d$. It is not hard to see that, up to reordering the associate matrices, the condition of being $P$-polynomial is equivalent to requiring that $p_{ij}^k = 0 $ whenever the sum of two of $\{i,j,k\}$ is strictly smaller than the third element. The latter condition is also called \textsl{metric} and we may use $P$-polynomial and metric exchangeably. The class metric schemes are exactly those where $A_1$ is the adjacency matrix of a distance regular graph. 

Similarly, we say a scheme is \textsl{$Q$-polynomial} or \textsl{cometric} if $E_i$ is a polynomial under Schur multiplication in $E_1$ of degree $i$, for each $i = 0,\ldots, d$. Equivalent, a scheme is $Q$-polynomial if $q_{ij}^k = 0 $ whenever the sum of two of $\{i,j,k\}$ is strictly smaller than the other element.


\subsection{Terwilliger algebra}\label{sec:Terwil}

We will now look at the Terwilliger algebra.
We consider an association scheme $\cA$ with associate matrices $A_0,\ldots, A_d$. Throughout this section, let $x\in V$ be a fixed vertex of $\cA$. For $i=0,\ldots, d$, we will define the diagonal matrix $E_i^*(x)$ as follows:
\[
E_i^*(x)_{y,y} = (A_i)_{x,y}.
\]
We call $E_i^*(x)$ the \textsl{$i$th dual idempotent with respect of $x$}.
Similarly, for $i =0,\ldots,d$, we will consider diagonal matrices $A_i^*(x)$ with entries as follows:
\[
A_i^*(x)_{y,y} = N (E_i)_{x,y}.
\]
We call $A_i^*$ the \textsl{$i$th dual distance matrix with respect to $x$}.
When the context is clear, we will write $E_i^*$ for $E_i^*(x)$ and $A_i^*$ for $A_i^*(x)$.

Note, that if $A_1$ is the adjacency matrix of a distance regular graph, $E_i^*$ is the diagonal characteristic matrix for the set of vertices at distance $i$ from $x$, also known as the \textsl{$i$th neighbourhood of $x$}. We also have that
\[
A_i^* A_j^* = \sum_{k=0}^d q_{ij}^k A_k^*
\]
from part (iv) of Theorem \ref{thm:drg}. 
The matrices $A_0^*,\ldots,A_d^*$ form a basis for some subspace $\cM^*$ of $\Mat{N}{N}{\cx}$. In $\cM^*$, the primitive idempotents are $E_0^*,\ldots,E_d^*$.

Recalling the eigenmatrices of a scheme, we can see that
\begin{equation}\label{eq:dualchangebases}
E_j^* =\frac{1}{N} \sum_{i = 0}^d P_{ij} A_i^* \text{ and } A_j^* =  \sum_{i=0}^d Q_{ij} E_i^*.
\end{equation}

The \textsl{Terwilliger algebra} $T(x)$ is the subalgebra of $\Mat{N}{N}{\cx}$ generated by $\cM$ and $\cM^*$. The following lemma is found in \cite{Ter92}. 

\begin{lemma}[Terwilliger 1992]\label{lem:Terwil92}
\[
\begin{split}
E_i^*A_j E_k^* = 0 &\Leftrightarrow p_{ij}^k =0, \text{ and; } \\
E_iA_j^* E_k = 0 &\Leftrightarrow q_{ij}^k =0. 
\end{split} 
\]\qed
\end{lemma}

The above lemma suggest that the Terwilliger algebra is easier to work with when many of the intersection numbers or the Krein parameters of the scheme vanish. We will later look at the class of Hadamard graphs, which have many vanishing Krein parameters and intersection numbers.

\section{Chopped correlation matrix for $P$- and $Q$-polynomials scheme}

Let $G$ be a distance regular graph of diameter $d$ that is $Q$-polynomial. We recall that, in this case $\delta=d$. We will consider the association scheme of $G$, with the idempotents in the $Q$-polynomial ordering. 
We choose as the matrix $A$ defining the Hamiltonian \eqref{eq:Hff2} the adjacency matrix $A_1$ of this scheme.
The eigenvalues of $A$ denoted previously by $\theta_i$ corresponds to
the constants $P_{i1}$ of the association scheme.

As seen previously, the matrices $A=A_1$ and $A^*=A_1^*$ can be expand in the basis of idempotents and dual idempotents, respectively: 
\begin{eqnarray}
 A&=&\sum_{i=0}^d P_{i,1} E_i\,, \label{eq:A1}\\
 A^*&=&\sum_{i=0}^d Q_{i,1} E_i^*\,.\label{eq:A1s}
\end{eqnarray}
However, due to the fact that we consider a graph in a $P-$ and $Q-$ polynomials scheme, $A$ and $A^*$ satisfy supplementary relations. Indeed, 
for a $P-$ and $Q-$ polynomials scheme, one gets $p_{1i}^j=0$ and $q_{1i}^j=0$ for $|i-j|>1$.
Then, from Lemma \ref{lem:Terwil92} and from $\sum_{i=0}^d E^*_i=1$, one deduces the following relation
\begin{eqnarray}
 A &=& \sum_{i=0}^d E_i^* A E_i^*+\sum_{i=1}^d\left( E_{i-1}^* A E_i^*  +E_{i}^* A E_{i-1}^* \right)\,.\label{eq:A1t}
 \end{eqnarray}
 Similarly, by using $\sum_{i=0}^d E_i=1$, one obtains
 \begin{eqnarray}
 A^* &=& \sum_{i=0}^d E_i A^* E_i+\sum_{i=1}^d\left( E_{i-1} A^* E_i  +E_{i} A^* E_{i-1} \right)\,.\label{eq:A1st}
\end{eqnarray}
In the basis where $E_i$ is diagonal, $A$ is diagonal and $A^*$ is tridiagonal by block. In contrary, in the basis where $E_i^*$ is diagonal, $A$ is tridiagonal by block and $A^*$ is diagonal by block. The choice of a graph $G$ in a $P-$ and $Q-$ polynomials association scheme leads to interesting properties for the matrix $A$ and $A^*$ which generalize the properties obtained from the fact to belong to a bispectral problem used in \cite{CNV1,CNV2}. Let us emphasize that $A$ and $A^*$ are not a tridiagonal pair \cite{IKT} in general  since there exists a subspace $\mathcal{W}$ of the space $\mathcal{V}$ spans by the vertices of the graph such that $A\mathcal{W}\subseteq \mathcal{W},A^*\mathcal{W}\subseteq \mathcal{W}$, $\mathcal{W}\neq 0$ and $\mathcal{W}\neq \mathcal{V}$. However restricted to that space $\mathcal{W}$, $A$ and $A^*$ become a tridiagonal pair.

Recall that we are concerned with the eigenvalues of the chopped correlation matrix
\[
C =\Pi(K,\ell)= \pi_1(\ell) \pi_2(K) \pi_1(\ell)\,,
\]
where we recall that $\pi_1(\ell)=\sum_{s=0}^\ell E_s^* $ and $\pi_2(K)=\sum_{k=0}^K E_k$.
Note that matrix $\Pi(j,\ell)$ is not in the Bose-Mesner algebra of $G$, but it is in the Terwilliger algebra of $G$.

\subsection{Properties of particular chopped correlation matrix}

We will state, without proof, some simple observations about $\Pi(K,\ell)$ in the following lemma. We denote by $\zeros$ a block of $0$s of the appropriate size. We will give the spectrum as a list of distinct eigenvalues $\theta$ with their multiplicities $m_{\theta}$ in superscript in parentheses. For example, if the eigenvalues of $M$ are $a$ with multiplicity $1$, $1$ with multiplicity $n$ and $0$ with multiplicity $3n-1$, we write the spectrum of M as $\{a^{(1)}, 1^{(n)}, 0^{(3n-1)}\}$. We denote by $\zeros_{m\times n}$ the $m\times n$ matrix of all zeroes and we omit the subscript when the order is clear from the context. 

\begin{lemma}\label{lem:pjell-easy} Let $\cA$ be an association scheme and $\Pi(K,\ell)$ defined as above. Let $N_k = \sum_{s=0}^k n_s$ and let $F_{k} = \sum_{s=0}^k f_s$. The following hold.
\begin{enumerate}[(i)]
\item $\Pi(d,d) = I$ and has spectrum $\{1^{(N)}\}$.
\item $\Pi(d,\ell) = \sum_{s=0}^{\ell} E_{s}$  and has spectrum $\{1^{(F_{\ell})}, 0^{(N-F_{\ell})}\}$.
\item $\Pi(j,d) = \pmat{I_{N_j} & \zeros \\ \zeros & \zeros }$  and has spectrum $\{1^{(N_j)},0^{(N-N_{j})} \}$.
\item $\Pi(0,0) = \pmat{\frac{1}{N} &\zeros \\ \zeros & \zeros }$  and has spectrum $\{\frac{1}{N}^{(1)}, 0^{(N-1)}\}$.
\item $\Pi(0,\ell) = \pmat{\frac{1}{N} F_{\ell} &\zeros \\ \zeros & \zeros }$  and has spectrum $\{\frac{F_{\ell}}{N}^{(1)}, 0^{(N-1)}\}$.
\item $\Pi(j,0) = \pmat{\frac{1}{N} \ones_{N_j\times N_j} &\zeros \\ \zeros & \zeros }$  and has spectrum $\{\frac{N_{j}}{N}^{(1)}, 0^{(N-1)}\}$.
\end{enumerate}
\end{lemma}

\subsection{(Generalized) Heun operator \label{sec:heun}}

For $\mu, \nu \in \re$, we define the \textsl{Heun operator} $T$ as follows:
\begin{eqnarray}
 T=\{A,A^* \} +\mu A^* +\nu A \ ,
\end{eqnarray}
where  $\{\cdot,\cdot\}$ denotes the anti-commutator bracket; that is, $\{A,B\} := AB + BA$ for matrices $A,B$. 
From relations \eqref{eq:A1}-\eqref{eq:A1st} and the properties of the idempotents, we obtain two expressions for the Heun operator:
\begin{eqnarray}
 T&=& \sum_{i=0}^d  \left( 2Q_{i,1}+\nu\right)  E_i^* A E_i^* + \mu \sum_{i=0}^d Q_{i,1} E_i^*\nonumber \\
 && + \sum_{i=1}^d \left( Q_{i-1,1}+ Q_{i,1}+ \nu\right)\left(E_{i-1}^* A E_i^* + E_{i}^* A E_{i-1}^*\right)\label{eq:T1}\\
 &=& \sum_{i=0}^d  \left(2P_{i,1}+\mu\right)  E_i A^* E_i + \nu \sum_{i=0}^d P_{i,1} E_i\nonumber \\
 && + \sum_{i=1}^d \left( P_{i-1,1}+ P_{i,1}+ \mu\right)\left(E_{i-1} A^* E_i + E_{i} A^* E_{i-1}\right)\,.\label{eq:T2}
\end{eqnarray}
We will give a choice of $\mu$ and $\nu$ such that the Heun operator commutes with the operator that we are considering. 
\begin{theorem} 
 For $\nu=-Q_{\ell,1}-Q_{\ell+1,1}$ and $\mu=-P_{K,1}-P_{K+1,1}$, these relations hold for the Heun operator
 \begin{equation}
  [T,\pi_1(\ell)]=0 \quad \text{and} \qquad  [T,\pi_2(K)]=0 \ .
 \end{equation}
\end{theorem}
\proof
By using \eqref{eq:T1}, one gets 
\begin{equation}
  [T,\pi_1(\ell)]=- (Q_{\ell,1} + Q_{\ell+1,1}+\nu)\left(E_{\ell}^* A E_{\ell+1}^* -E_{\ell+1}^* A E_{\ell}^*\right)\,.
\end{equation}
With the value of $\nu$ given in the theorem the commutator vanishes. The second relation is proven similarly by using \eqref{eq:T2}. \qed 

Let $T(K,\ell)$ denote the Heun operator with  $\nu=-Q_{\ell,1}-Q_{\ell+1,1}$ and $\mu=-P_{K,1}-P_{K+1,1}$.

\begin{corollary}\label{cor:Tcommutes} $T(K,\ell)$  commutes with $\Pi(K,\ell)$. \end{corollary}

\subsection{Formally self-dual association schemes}

We say that an association scheme is \textsl{formally self-dual} if, for some ordering of the primitive idempotents, $P=Q$. It follows then that the intersection numbers and Krein parameters agree; more formally, for all $0 \leq i,j,k \leq d$,
\[
p_{ij}^k = q_{ij}^k.
\]
The Hamming scheme and the Hadamard scheme are examples of formally self-dual schemes. 

Let $\cA$ be a formally self-dual, metric and co-metric association scheme. We may assume that the primitive idempotents $\{E_i\}_{i=0}^d$, are ordered such that $n_i = f_i$. With this ordering, the $P$- and $Q$ matrices of the scheme are equal. In particular $Q_{i,1}=\theta_i$.

\begin{theorem} \label{th:dual}
If $G$ is a graph whose adjacency matrix is the $A_1$ matrix of a formally self-dual, metric and co-metric association scheme,  $\Pi(K,\ell)$ is cospectral to $\Pi(\ell,K)$ for all $K,\ell = 0,\ldots,d$.
\end{theorem}

\proof We have shown previously that $D$ defined by \eqref{eq:defD} is cospectral with $C = \Pi(\ell,K)$.
Now we show $D$ is cospectral to $\Pi(K,\ell)$.  \qed 

For bipartite graph in a $Q$-polynomial and $Q$-bipartite scheme, one gets 
\begin{equation}
 E_i^* A E_i^*=0 \quad \text{and} \qquad  E_i A^* E_i=0 \,,
\end{equation}
for all $i =0 , \ldots, d$.
By taking the forms \eqref{eq:A1} and \eqref{eq:A1st} for $A$ and $A^*$, one can show that
\begin{equation}
 A^2A^*-\rho AA^*A+A^*A^2-\tau A^*=\sum_{i=1}^d(\theta_i^2-\rho\theta_i\theta_{i-1}+\theta_{i-1}^2-\tau)\left( E_{i-1} A^* E_i  +E_{i} A^* E_{i-1} \right)\,,
\end{equation}
where we recall that $\theta_i=P_{i,1}$. 
Similarly, by taking the forms \eqref{eq:A1s} and \eqref{eq:A1t} for $A$ and $A^*$, one can show that
\begin{equation}
 (A^*)^2A-\rho A^*AA^*+A(A^*)^2-\tau A=\sum_{i=1}^d(\theta_i^2-\rho\theta_i\theta_{i-1}+\theta_{i-1}^2-\tau)\left( E_{i-1} A E_i  +E_{i} A E_{i-1} \right)\,,
\end{equation}
where we recall that $Q_{i,1}=\theta_i$ for formally self-dual association schemes.

The Hadamard graph of order $n$ (see below for the definition) is bipartite and formally self-dual with eigenvalues $\theta_0=n, \theta_1=\sqrt{n}, \theta_2=0, \theta_3=-\sqrt{n}, \theta_4=-n$.
Then, for $\tau=n$ and $\rho=\sqrt{n}$, one gets
$\theta_i^2-\rho\theta_i\theta_{i-1}+\theta_{i-1}^2-\tau=0$ for $i=1,2,3,4$. One gets 
\begin{eqnarray}
 && A^2A^*-\sqrt{n} AA^*A+A^*A^2-n A^*=0\,,\\
 && (A^*)^2A-\sqrt{n} A^*AA^*+A(A^*)^2-n A= 0\,.
\end{eqnarray}

For Hamming graph, one gets $\theta_i=-L+2i$ (for $i=0,1,\dots L$). For $\tau=4$ and $\rho=2$, one gets 
$\theta_i^2-\rho\theta_i\theta_{i-1}+\theta_{i-1}^2-\tau=0$ and we recover the result \cite{GO}:
\begin{eqnarray}
 && A^2A^*-2 AA^*A+A^*A^2-4 A^*=0\,,\\
 && (A^*)^2A-2 A^*AA^*+A(A^*)^2-4 A= 0\,.
\end{eqnarray}

\section{Hadamard graphs}\label{sec:hadamardgrsinfo}

\subsection{Definition of the Hadamard graphs}

An example of a formally self-dual association scheme is the Bose-Mesner scheme of the Hadamard graphs. Hadamard graphs are exposited in Section 1.8 of \cite{BCN}. We will give a full description here. The Hadamard graph of order $n$ is a distance-regular, antipodal and bipartite graphs constructed as follows. Let $H$ be an $n\times n$ Hadamard matrix. The graph $\cH$ of $H$ has two vertices $c^+$ and $c^-$ for each column of $H$ and vertices $r^+$ and $r^-$ for each row of $H$. For $0\leq i,j\leq n-1$, the graph $\cH$ has edges $(r_i^+, c_j^+)$ and  $(r_i^-, c_j^-)$ if $H_{i,j} =1$ and edges $(r_i^+, c_j^-)$ and  $(r_i^-, c_j^+)$ if $H_{i,j} =-1$. The resulting graph has diameter $4$ and $4n$ vertices. For the ordering of the idempotents as above, the $P$ and $Q$ matrices are equal and are as follows:
\[
P = Q = \pmat{ 1 & n & 2n- 2 & n & 1 \\
1 & \sqrt{n} & 0 & -\sqrt{n} & -1 \\
1 & 0 & -2 & 0 & 1 \\
1 & -\sqrt{n} & 0 & \sqrt{n} & -1\\
1 & -n & 2n- 2 & -n & 1 \\}.
\]
An equivalent definition for a distance-regular graph is a graph for which there exist integers $\{b_i\}_{i=0}^{d-1} , \{c_i\}_{i=1}^d$ such that for any two vertices $x,y$ in G at distance $i$, there are exactly $c_i$ neighbours of y at distance $i-1$ from $x$ and exactly $b_i$ neighbors of $y$ at distance $i+1$ from $x$. The array $\{b_0,\ldots, b_{d-1}: c_1,\ldots, c_d\}$ is the \textsl{intersection array} of the graph. The intersection array of the Hadamard graph is $\{n, n-1 ,\frac{n}{2},1: 1, \frac{n}{2}, n-1, n\}$. 
Note that since the Hadamard graph is bipartite, $p_{ij}^k = 0$ whenever $i+j+k$ is odd.

\begin{figure}
    \centering
    \includegraphics{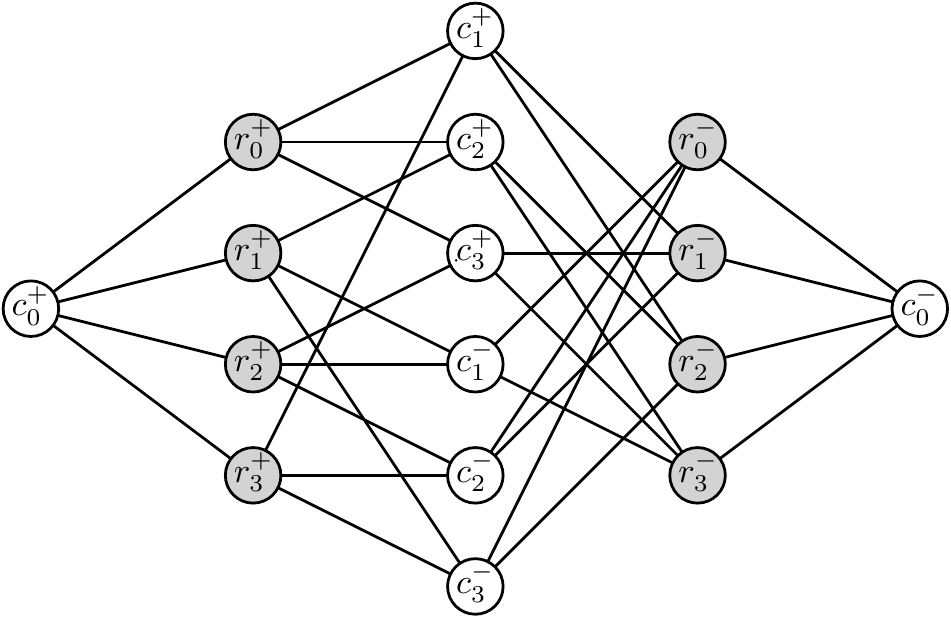}
    \caption{The Hadamard graph, $\cH_4$, of the $4\times 4$ Hadamard matrix $H_4$. \label{fig:had4}}
\end{figure}

The unique Hadamard graph of order $2$ is the cycle on $8$ vertices. Up to permutation of row and columns, multiplying rows and columns by $-1$ and taking transposes, all Hadamard matrices of order $4$ are equivalent to the following: 
\[
H_4 = \pmat{1 & 1 & 1 & 1 \\
           1 & -1 &1 & -1 \\
           1 & -1 & -1 & 1\\
           1 & 1 & -1 &-1 }.
\]
Figure \ref{fig:had4} shows the Hadamard graph $\cH_4$ corresponding to $H_4$, with its two color classes shown in white and gray. Up to isomorphism, $\cH_4$ is the unique Hadamard graph of order $4$, on $16$ vertices. We observe that $\cH_4$ is isomorphic to the hypercube graph $Q_4$; note that no other Hadamard graph is isomorphic to the hypercube on the same number of vertices, since the graphs will have different diameters. 

Let $H$ be Hadamard matrix of $n$. We may assume that the first row and first column of $H$ all have entry $1$. Let $\comp{H}$ be obtained from $H$ by deleting the first column. We consider the Hadamard graph constructed from $H$ and we will write the distance matrices as block matrices, partitioned by the distance partition with respect to vertex $0$. The block sizes are $1$, $n$, $2n-2$, $n$ and $1$ and the sets are:
\[
\{c_0^+\}, \{r_i^+\}_{i=0}^{n-1},\{c_1^+,\ldots,c_{n-1}^+,c_1^-,\ldots,c_{n-1}^-\} , \{r_i^-\}_{i=0}^{n-1},\{c_0^-\}
\]
with the ordering of the vertices as given above.

Let $R_m$ be $m\times m$ back diagonal matrix given by $R_{i,m+1-i} = 1$ for $i=1,\ldots, m$ and all other entries are $0$. Note that  $\zeros_{n\times m}$ and $J_{n\times m}$ denote the $n\times m$ block of $0$s and $1$s, respectively. We denote by $\ones_m$ a column vector of order $m$ with all entries equal to $1$. Recall that $I_m$ is the identity matrix of order $m$. Note that we will write our matrices as block matrices and the blocks are separated by lines. When the order of the submatrix is clear, we will omit the subscript of $\zeros_{n\times m}$, $J_{n\times m}$, $\ones_m$ and $I_m$.  We let
\[M_1 = \frac{1}{2} \left( \begin{array}{c|c} J_{n,n-1} + \comp{H} & J_{n,n-1} - \comp{H}\end{array} \right)
\]
 and
 \[M_2 = \frac{1}{2}  \left( \begin{array}{c|c} J_{n,n-1} - \comp{H} & J_{n,n-1} + \comp{H}\end{array} \right).
\]
Note that $M_1 +M_2= J_{n, 2n-2}$ and $M_1 -M_2 =  \left( \begin{array}{c|c} J_{n,n-1} & -J_{n,n-1}\end{array} \right)$. We observe that $\comp{H}\comp{H}^T  = n I_n - J_n$, so we may compute
\[
M_1 M_1^T = M_2M_2^T = \frac{1}{2}(nI_n + (n-2) J_n)
\]
and
\[ M_2 M_1^T = M_1 M_2^T = \frac{n}{2}(J_n - I_n).
\]
With this setup, we may write the distance matrices more explicitly as block matrices:
\begin{equation}\label{eq:HadA1A3}  A_1 = \left( \begin{array}{c|c|c|c|c}	0 & \ones_n^T & \zeros & \zeros & \zeros \\
      \hline
			\ones_n & \zeros & M_1 & \zeros & \zeros \\
      \hline
			\zeros & M_1^T & \zeros & M_2^T & \zeros \\
      \hline
			\zeros & \zeros & M_2 & \zeros & \ones_n \\
      \hline
			\zeros & \zeros & \zeros & \ones_n^T & 0 \end{array} \right)
\text{ and }
 A_3 = \left( \begin{array}{c|c|c|c|c}	0 & \zeros & \zeros &  \ones_n^T & \zeros \\
      \hline
				\zeros & \zeros & M_2 & \zeros & \ones_n  \\
              \hline
				\zeros & M_2^T & \zeros & M_1^T & \zeros \\
              \hline
				\ones_n & \zeros & M_1 & \zeros & \zeros  \\
              \hline
				\zeros & \ones^T & \zeros & \zeros & 0 \end{array} \right).
\end{equation}
 Let $R_m$ be the $m\times m$ back-diagonal matrix. Then
 \vspace{10pt} 
 
\begin{equation}\label{eq:HadA2A4} A_4 =  \left( \begin{array}{c|c|c|c|c}	0 & \zeros & \zeros &  \zeros & 1 \\
       \hline
 				\zeros & \zeros & \zeros  & I_{n} & \zeros   \\
              \hline
 				\zeros & \zeros  & R_2\otimes I_{n-1} & \zeros  & \zeros \\
              \hline
 				\zeros  & I_n & \zeros  & \zeros & \zeros  \\
              \hline
 				1 & \zeros & \zeros & \zeros & 0 \end{array} \right)
\text{and }
 A_2 \!=\! \left( \begin{array}{c|c|c|c|c}	0 & \zeros & \ones_{2n-2}^T & \zeros & \zeros \\
      \hline
				\zeros & J_n - I_n & \zeros & J_n - I_n & \zeros \\
              \hline
				\ones_{2n-2} & \zeros & J_{2n-2}\!-\!J_2\!\otimes\! I_{n-1} &\zeros & \ones \\
              \hline
				\zeros & J_n - I_n & \zeros & J_n- I_n & \zeros \\
              \hline
				\zeros & \zeros & \ones^T & \zeros & 0 \end{array} \right)\!.
\end{equation}

 \vspace{10pt} 
\noindent For completeness, $A_0 = I_{4n}$.

\subsection{Direct computations of the spectrum for chopped correlation matrices}

We will find simpler expression for $\Pi(j,\ell)$ for a Hadamard graph of order $n$ using \eqref{eq:HadA1A3} and \eqref{eq:HadA2A4}. In light of Lemma \ref{lem:pjell-easy}, we will focus on $\Pi(j,\ell)$ for $j,\ell \in \{1,2,3\}$.

\begin{lemma}\label{lem:Pi-Had-explicit} For any $n$, we consider $\Pi(j,\ell)$ with respect to vertex $0$. The following hold.
\begin{enumerate}[(i)]
\item $\Pi(1,1)= \frac{1}{4n} \pmat{n+1 & (1+\sqrt{n})\ones_n^T & \zeros \\
								(1+\sqrt{n})\ones_n & J_n + n I_n & \zeros \\
								\zeros & \zeros & \zeros}$.
\item \resizebox{\textwidth}{!}{$\Pi(2,1) = \frac{1}{4n}\!\left( \begin{array}{c|c|c|c}	n + 1 & (\sqrt{n}+1) \ones_n^T & \ones_{2n-2}^T &  \zeros \\
            \hline
      			(\sqrt{n}+1) \ones_n  & J_n+ n I_n &  \left( \begin{array}{c|c}  J_{n,n-1}+ \sqrt{n}\comp{H} &  J_{n,n-1} - \sqrt{n}\comp{H}\end{array} \right) & \zeros \\
            \hline
      			\ones_{2n-2} &  \left( \begin{array}{c}  J_{n-1,n}+ \sqrt{n}\comp{H}^T \\ \hline  J_{n-1,n} - \sqrt{n}\comp{H}^T\end{array} \right) & \pmat{ J_{n-1} + nI_{n-1} &  J_{n-1} -nI_{n-1} \\  J_{n-1} -nI_{n-1} &  J_{n-1} + nI_{n-1}} & \zeros \\
            \hline
      			\zeros & \zeros & \zeros & \zeros
              \end{array} \right)$
      }
\item  \resizebox{\textwidth}{!}{$\Pi(3,1)=\! \frac{1}{4n}\!\left( \begin{array}{c|c|c|c|c}	n + 1 & (\sqrt{n}+1) \ones_n^T & \ones_{2n-2}^T & ( 1- \sqrt{n}) \ones_n^T  & 0 \\
            \hline
      			(\sqrt{n}+1) \ones_n  & J_n+ n I_n &  \left( \begin{array}{c|c}  J_{n,n-1}+ \sqrt{n}\comp{H} &  J_{n,n-1} - \sqrt{n}\comp{H}\end{array} \right) & J_n -n I_n & \zeros_n \\
            \hline
      			\ones_{2n-2} &  \left( \begin{array}{c}  J_{n-1,n}+ \sqrt{n}\comp{H}^T \\ \hline  J_{n-1,n} - \sqrt{n}\comp{H}^T\end{array} \right) & \pmat{ J_{n-1} + nI_{n-1} &  J_{n-1} -nI_{n-1} \\  J_{n-1} -nI_{n-1} &  J_{n-1} + nI_{n-1}} & \left( \begin{array}{c}  J_{n-1,n}+ \sqrt{n}\comp{H}^T \\ \hline  J_{n-1,n} - \sqrt{n}\comp{H}^T\end{array} \right) & \zeros_{2n-2} \\
            \hline
      			( 1-\sqrt{n}) \ones_n & J_n -n I_n &  \left( \begin{array}{c|c}  J_{n,n-1}+ \sqrt{n}\comp{H} &  J_{n,n-1} - \sqrt{n}\comp{H}\end{array} \right) & J_n +n I_n & \zeros_n \\
            \hline
      			0 & \zeros_n^T & \zeros_{2n-2}^T & \zeros_n^T & 0
              \end{array} \right)$
          }
\item $\Pi(1,2)= \frac{1}{4n} \pmat{3n-1 & (1+\sqrt{n})\ones_n^T & \zeros \\
								(1+\sqrt{n})\ones_n & 3nI_n - J_n & \zeros \\
								\zeros & \zeros & \zeros}$.
\item \resizebox{\textwidth}{!}{$\Pi(2,2) = \frac{1}{4n}\!\left( \begin{array}{c|c|c|c}	3n - 1 & (\sqrt{n}+1) \ones_n^T & - \ones_{2n-2}^T &  \zeros \\
            \hline
      			(\sqrt{n}+1) \ones_n  &  3n I_n - J_n &  \left( \begin{array}{c|c}  J_{n,n-1}+ \sqrt{n}\comp{H} &  J_{n,n-1} - \sqrt{n}\comp{H}\end{array} \right) & \zeros \\
            \hline
      			- \ones_{2n-2} &  \left( \begin{array}{c}  J_{n-1,n}+ \sqrt{n}\comp{H}^T \\ \hline  J_{n-1,n} - \sqrt{n}\comp{H}^T\end{array} \right) &  \pmat{\scriptstyle 3nI_{n-1} - J_{n-1} & \scriptstyle nI_{n-1} - J_{n-1} \\ \scriptstyle nI_{n-1}-J_{n-1} & \scriptstyle 3nI_{n-1}-J_{n-1}} & \zeros \\
            \hline
      			\zeros & \zeros & \zeros & \zeros
              \end{array} \right)$
      }
\item \noindent\resizebox{\textwidth}{!}{
$
\Pi(3,2) 
=\! \scriptstyle\frac{1}{4n}\!\left( \begin{array}{c|c|c|c|c}
      3n - 1 & (\scriptstyle\sqrt{n}+1) \ones_n^T & -\ones_{2n-2}^T & (\scriptstyle 1- \sqrt{n}) \ones_n^T  & 0 \\
      \hline
			(\scriptstyle\sqrt{n}+1) \ones_n  & 3n I_n - J_n  & \scriptstyle \left( \begin{array}{c|c} \scriptstyle J_{n,n-1}+ \sqrt{n}\comp{H} & \scriptstyle J_{n,n-1} - \sqrt{n}\comp{H}\end{array} \right) & n I_n - J_n & \zeros \\
      \hline
			-\ones_{2n-2} & \scriptstyle \left( \begin{array}{c} \scriptstyle J_{n-1,n}+ \sqrt{n}\comp{H}^T \\ \hline \scriptstyle J_{n-1,n} - \sqrt{n}\comp{H}^T\end{array} \right) &\scriptstyle \pmat{\scriptstyle 3nI_{n-1} - J_{n-1} & \scriptstyle nI_{n-1} - J_{n-1} \\ \scriptstyle nI_{n-1}-J_{n-1} & \scriptstyle 3nI_{n-1}-J_{n-1}} &\scriptstyle \left( \begin{array}{c} \scriptstyle J_{n-1,n}+ \sqrt{n}\comp{H}^T \\ \hline \scriptstyle J_{n-1,n} - \sqrt{n}\comp{H}^T\end{array} \right) & \zeros \\
      \hline
			(\scriptstyle 1-\sqrt{n}) \ones_n & n I_n - J_n & \scriptstyle \left( \begin{array}{c|c} \scriptstyle J_{n,n-1}+ \sqrt{n}\comp{H} & \scriptstyle J_{n,n-1} - \sqrt{n}\comp{H}\end{array} \right) & 3n I_n - J_n &  \zeros \\
      \hline
		0 & \zeros_n^T & \zeros_{2n-2}^T & \zeros_n^T & 0
        \end{array} \right).
$
}
\item $\Pi(1,3)= \pmat{I_{n+1} & \zeros \\ \zeros & \zeros} - \frac{1}{4n} \pmat{1 & -\ones_n^T & \zeros \\
								-\ones_n & J_n & \zeros \\
								\zeros & \zeros & \zeros}$.
\item $\Pi(2,3)= \pmat{I_{3n-1} & \zeros \\ \zeros & \zeros} - \frac{1}{4n} \pmat{1 & -\ones_n^T & \ones_{2n-2}^T & \zeros \\
								-\ones_n & J_n &  - J_{n,2n-2} & \zeros \\ 
								\ones_{2n-2} & - J_{2n-2,n} & J_{2n-2} & \zeros \\
								\zeros & \zeros & \zeros & \zeros }$.
\item $\Pi(3,3)= \pmat{I_{4n-1} & \zeros \\ \zeros & \zeros} - \frac{1}{4n} \pmat{1 & -\ones_n^T & \ones_{2n-2}^T & - \ones_n^T & 0 \\
								-\ones_n & J_n &  - J_{n,2n-2} & J_n & \zeros \\ 
								\ones_{2n-2} & - J_{2n-2,n} & J_{2n-2} & -J_{2n-2,n} & \zeros \\
								-\ones_n & J_n &  - J_{n,2n-2} & J_n & \zeros \\ 
								0 & \zeros & \zeros & \zeros & 0 }$.
\end{enumerate}
\end{lemma}
\proof We will start with $\ell =1$. The projector $\Pi(j,1)$ is the square, order $\sum_{i=0}^j n_j$, principal submatrix of $E_0 + E_1$.  Recall that $M_1 +M_2= J_{n, 2n-2}$ and $M_1 -M_2 =  \left( \begin{array}{c|c} \comp{H} & -\comp{H}\end{array} \right)$. 
We have that
\[E_0 + E_1 = \frac{1}{4n} J_{4n} + \frac{1}{4n}\left(nA_0 + \sqrt{n}A_1 - \sqrt{n}A_3 - n A_4\right) \]
and so

\noindent\resizebox{\textwidth}{!}{
$
E_0 + E_1
=\! \scriptstyle\frac{1}{4n}\!\left( \begin{array}{c|c|c|c|c}	n + 1 & (\scriptstyle\sqrt{n}+1) \ones_n^T & \ones_{2n-2}^T & (\scriptstyle 1- \sqrt{n}) \ones_n^T  & 1-n \\
      \hline
			(\scriptstyle\sqrt{n}+1) \ones_n  & J_n+ n I_n & \scriptstyle \left( \begin{array}{c|c} \scriptstyle J_{n,n-1}+ \sqrt{n}\comp{H} & \scriptstyle J_{n,n-1} - \sqrt{n}\comp{H}\end{array} \right) & J_n -n I_n & (\scriptstyle1- \sqrt{n}) \ones_n \\
      \hline
			\ones_{2n-2} & \scriptstyle \left( \begin{array}{c} \scriptstyle J_{n-1,n}+ \sqrt{n}\comp{H}^T \\ \hline \scriptstyle J_{n-1,n} - \sqrt{n}\comp{H}^T\end{array} \right) &\scriptstyle \pmat{\scriptstyle J_{n-1} + nI_{n-1} & \scriptstyle J_{n-1} -nI_{n-1} \\ \scriptstyle J_{n-1} -nI_{n-1} & \scriptstyle J_{n-1} + nI_{n-1}} &\scriptstyle \left( \begin{array}{c} \scriptstyle J_{n-1,n}+ \sqrt{n}\comp{H}^T \\ \hline \scriptstyle J_{n-1,n} - \sqrt{n}\comp{H}^T\end{array} \right) & \ones_{2n-2} \\
      \hline
			(\scriptstyle 1-\sqrt{n}) \ones_n & J_n -n I_n & \scriptstyle \left( \begin{array}{c|c} \scriptstyle J_{n,n-1}+ \sqrt{n}\comp{H} & \scriptstyle J_{n,n-1} - \sqrt{n}\comp{H}\end{array} \right) & J_n +n I_n &  (\scriptstyle \sqrt{n}+1) \ones_n \\
      \hline
			1-n & (\scriptstyle 1- \sqrt{n}) \ones_n^T & \ones_{2n-2}^T & (\scriptstyle \sqrt{n} +1) \ones_n^T & n+1
        \end{array} \right).
$
}
This gi4nes us the expressions for $\Pi(1,1)$, $\Pi(2,1)$ and $\Pi(3,1)$.

We have that
\[E_0 + E_1 + E_2 = (E_0 + E_1) + (2n-2)I - 2A_2 + (2n-2) A_4\]
and so

\noindent\resizebox{\textwidth}{!}{
$
E_0 + E_1 + E_2
=\! \scriptstyle\frac{1}{4n}\!\left( \begin{array}{c|c|c|c|c}
      3n - 1 & (\scriptstyle\sqrt{n}+1) \ones_n^T & -\ones_{2n-2}^T & (\scriptstyle 1- \sqrt{n}) \ones_n^T  & n-3 \\
      \hline
			(\scriptstyle\sqrt{n}+1) \ones_n  & 3n I_n - J_n  & \scriptstyle \left( \begin{array}{c|c} \scriptstyle J_{n,n-1}+ \sqrt{n}\comp{H} & \scriptstyle J_{n,n-1} - \sqrt{n}\comp{H}\end{array} \right) & n I_n - J_n & (\scriptstyle1- \sqrt{n}) \ones_n \\
      \hline
			-\ones_{2n-2} & \scriptstyle \left( \begin{array}{c} \scriptstyle J_{n-1,n}+ \sqrt{n}\comp{H}^T \\ \hline \scriptstyle J_{n-1,n} - \sqrt{n}\comp{H}^T\end{array} \right) &\scriptstyle \pmat{\scriptstyle 3nI_{n-1} - J_{n-1} & \scriptstyle nI_{n-1} - J_{n-1} \\ \scriptstyle nI_{n-1}-J_{n-1} & \scriptstyle 3nI_{n-1}-J_{n-1}} &\scriptstyle \left( \begin{array}{c} \scriptstyle J_{n-1,n}+ \sqrt{n}\comp{H}^T \\ \hline \scriptstyle J_{n-1,n} - \sqrt{n}\comp{H}^T\end{array} \right) & -\ones_{2n-2} \\
      \hline
			(\scriptstyle 1-\sqrt{n}) \ones_n & n I_n - J_n & \scriptstyle \left( \begin{array}{c|c} \scriptstyle J_{n,n-1}+ \sqrt{n}\comp{H} & \scriptstyle J_{n,n-1} - \sqrt{n}\comp{H}\end{array} \right) & 3n I_n - J_n &  (\scriptstyle \sqrt{n}+1) \ones_n \\
      \hline
			n-3 & (\scriptstyle 1- \sqrt{n}) \ones_n^T & -\ones_{2n-2}^T & (\scriptstyle \sqrt{n} +1) \ones_n^T & 3n-1
        \end{array} \right).
$
}
This gives us the expressions for $\Pi(1,2)$, $\Pi(2,2)$ and $\Pi(3,2)$.

We have that 
\[
E_0 + E_1 + E_2 + E_3 = I - E_4 = I - \frac{1}{4n} 
\left(\begin{array}{c|c|c|c|c} 
    1 & -\ones_{n}^T & \ones_{2n-2}^T & - \ones_{n}^T & 1 \\
\hline 
    - \ones_{n} & J_n & -J_{n, 2n-2} & J_n & -\ones_{n} \\
\hline
    \ones_{2n-2} & -J_{2n-2,n} & J_{2n-2} & -J_{2n-2,n} & \ones_{2n-2} \\
\hline
     - \ones_{n} & J_n & -J_{n, 2n-2} & J_n & -\ones_{n} \\
\hline
     1 & -\ones_{n}^T & \ones_{2n-2}^T & - \ones_{n}^T & 1 \\
\end{array} \right)
\]
from whence we obtain the expressions for  $\Pi(1,3)$, $\Pi(2,3)$ and $\Pi(3,3)$. \qed 

Given the expressions in Lemma \ref{lem:Pi-Had-explicit}, a direct computation permits to explicitly find the eigenvalues of  $\Pi(1,3)$, $\Pi(2,3)$, $\Pi(3,3)$, $\Pi(1,1)$ and $\Pi(1,2)$ given in the next proposition. The fact that $\Pi(j,\ell)$ is cospectral to $\Pi(\ell,j)$ provides the eigenvalues for other Hadamard graphs. The more complicated computation of the spectrum of $\Pi(2,2)$ is the subject of the next subsection.
Recall that we write the spectrum of a matrix $M$
as a set of its distinct eigenvalues $\theta$ with their multiplicities $m_{\theta}$ in superscript in round brackets. 

\begin{proposition}\label{lem1}
For $G$ a Hadamard grapf of order $n$, we consider $\Pi(j,\ell)$ as above. The following are true:
\begin{enumerate}[(i)]
    \item The spectrum of $\Pi(1,3)$ and $\Pi(3,1)$ is  $\{ \left(\frac{3n-1}{4n}\right)^{(1)}, 1^{(n)}, 0^{(3n-1)} \}$.
    \item The spectrum of $\Pi(2,3)$ and $\Pi(3,2)$ is $\{ \left(\frac{n+1}{4n}\right)^{(1)}, 1^{(3n-2)}, 0^{(n+1)} \}$.
    \item The spectrum of $\Pi(3,3)$ is $\{ \left(\frac{1}{4n}\right)^{(1)}, 1^{(4n-2)}, 0^{(1)} \}$.
    \item The spectrum of $\Pi(1,1)$ is  \[\left\{0^{(3n-1)},\, \nicefrac{1}{4}^{(n-1)},\, \frac{2n + 5- \sqrt{16n + 32\sqrt{n} + 25}}{8n}^{(1)},\, \frac{2n + 5+ \sqrt{16n + 32\sqrt{n} + 25}}{8n}^{(1)}     \right\}.\]
    \item The spectrum of $\Pi(1,2)$ and $\Pi(2,1)$ is  \[\left\{0^{(3n-1)},\, \nicefrac{3}{4}^{(n-1)},\, \frac{6n - 5+ \sqrt{16n + 32\sqrt{n} + 25}}{8n}^{(1)},\, \frac{6n - 5 - \sqrt{16n + 32\sqrt{n} + 25}}{8n}^{(1)}     \right\}.\]
\end{enumerate}
\end{proposition}

\proof By using Theorem \ref{th:dual}, the spectrum of $\Pi(j,\ell)$ for $j>\ell$ is similar to the ones of $\Pi(\ell,j)$. We need to prove the above theorem only for $\Pi(j,\ell)$ with $j\leq \ell$. 

Observe that $\Pi(1,3)$ has only non-zero $(n+1)\times (n+1)$ block. Thus the eigenvalues of $\Pi(1,3)$ are $0^{(3n-1)}$ and the eigenvalues of $I_{n+1} - \frac{1}{4n}N_1$, where $N_1$ is the first  $(n+1)\times (n+1)$ principal submatrix of $4nE_4$. We observe that $E_4$ has rank $1$ and so $N_1$ has only one non-zero eigenvalue. We see that 
\[ N_1 \pmat{1\\ -\ones_n} = (n+1) \pmat{1\\ -\ones_n}.
\]
Let $\Zv$ and $\lambda$ be such that $N_1\Zv = \lambda \Zv$. Then 
\[
\left( I_{n+1} - \frac{1}{4n}N_1\right) \Zv = \left( 1 - \frac{n+1}{4n} \right) \Zv. 
\]
Thus the spectrum of  $\Pi(1,3)$  is as follows: $0^{(3n-1)}$, $ \left( 1 - \nicefrac{n+1}{4n} \right)^{(1)}$ and $1^{(n)}$.

Similarly, the eigenvalues of $\Pi(2,3)$ are $0$ with multiplicity $n+1$ and the eigenvalues of  $I_{3n-1} - \frac{1}{4n}N_2$, where $N_1$ is the first  $(3n-1)\times (3n-1)$ principal submatrix of $4nE_4$. Similarly to the eigenvector for $N_1$, we see that $\pmat{1& -\ones_n^T & \ones_{2n-2}}^T$ is an eigenvector of $N_2$ with eigenvalues $3n-1$. Thus the spectrum of  $\Pi(1,3)$  is as follows: $0^{(n+1)}$, $ \left( 1 - \nicefrac{3n-1}{4n} \right)^{(1)}$ and $1^{(3n-2)}$.

Similarly, the eigenvalues of $\Pi(3,3)$ are $0$ with multiplicity $1$ and the eigenvalues of  $I_{3n-1} - \frac{1}{4n}N_2$, where $N_3$ is the first  $(4n-1)\times (4n-1)$ principal submatrix of $4nE_4$. Similarly to the eigenvector for $N_1$ and $N_2$, we see that $\pmat{1& -\ones_n^T & \ones_{2n-2} & -\ones_n^T}^T$ is an eigenvector of $N_3$ with eigenvalues $4n-1$. Thus the spectrum of  $\Pi(1,3)$  is as follows: $0^{(1)}$, $ \left( 1 - \nicefrac{4n-1}{4n} \right)^{(1)}$ and $1^{(4n-2)}$.

We now consider $\Pi(1,1)$. Let 
\[N_4 =\pmat{1 & (1+\sqrt{n})\ones_n^T  \\
								(1+\sqrt{n})\ones_n & J_n }. \]
The eigenvalues of $\Pi(1,1)$ are $0$ with multiplicity $3n-1$ and the eigenvalues of 
\[
 \frac{1}{4n} \pmat{n+1 & (1+\sqrt{n})\ones_n^T  \\
								(1+\sqrt{n})\ones_n & J_n + n I_n  }
  = \frac{1}{4n} \left( nI_{n+1} +  N_4\right).
\]
Observe that $N_4$ has rank $2$ and thus has eigenvalue $0$ with multiplicity $n-1$. For the two non-zero eigenvalues of $N_4$, observe the following:
\[
N_4 \pmat{1 \\ t \ones_n} = \pmat{1 + (1+\sqrt{n})tn \\ (1+\sqrt{n}+ tn) \ones_n} = (1 + (1+\sqrt{n})tn)  \pmat{1 \\ \frac{1+\sqrt{n}+ tn}{1 + (1+\sqrt{n})tn} \ones_n}.
\]
Consider the following:
\begin{align}
 t &= \frac{1+\sqrt{n}+ tn}{1 + (1+\sqrt{n})tn} \nonumber \\
 t + (1+\sqrt{n})t^2n &= 1+\sqrt{n}+ tn \nonumber \\
 (n + n\sqrt{n})t^2 + (1-n) - 1 - \sqrt{n} &= 0 \label{eq:n4tsoln} 
\end{align}
For the two solutions of \eqref{eq:n4tsoln}, we have that $\pmat{1 & t \ones_n^T}^T$ is an eigenvector of $N_4$ with eigenvalue $1 + (1+\sqrt{n})tn$. We apply the quadratic formula to obtain that these two eigenvalues are 
\[
\lambda_1 = \frac{5- \sqrt{16n + 32\sqrt{n} + 25}}{2}   \text{ and } \lambda_2 = \frac{5+ \sqrt{16n + 32\sqrt{n} + 25}}{2}.
\]
We have obtained that the spectrum of $\Pi(1,1)$ is
$\left\{0^{(3n-1)},\, \nicefrac{1}{4}^{(n-1)},\, \nicefrac{n + \lambda_1}{4n}^{(1)},\, \nicefrac{n + \lambda_2}{4n}^{(1)}     \right\}$.

We now consider $\Pi(1,2)$. Let 
\[N_5 =\pmat{1 & -(1+\sqrt{n})\ones_n^T  \\
								-(1+\sqrt{n})\ones_n & J_n }. \]
The eigenvalues of $\Pi(1,2)$ are $0$ with multiplicity $3n-1$ and the eigenvalues of 
\[
\frac{1}{4n} \pmat{3n-1 & (1+\sqrt{n})\ones_n^T & \zeros \\
								(1+\sqrt{n})\ones_n & 3nI_n - J_n & \zeros \\
								\zeros & \zeros & \zeros}
  = \frac{1}{4n} \left( 3nI_{n+1} - N_5\right).
\]
Observe that 
\[
\pmat{1 & \zeros^T \\ \zeros & -I_{n}} N_5 \pmat{1 & \zeros^T \\ \zeros & -I_{n}} = \pmat{1 & (1+\sqrt{n})\ones_n^T  \\
								(1+\sqrt{n})\ones_n & J_n } = N_4.
\]
Thus $N_5$ has the same eigenvalues as $N_4$, namely $\lambda_1$ and $\lambda_2$. This gives that that the spectrum of $\Pi(1,2)$ is
$\left\{0^{(3n-1)},\, \nicefrac{3}{4}^{(n-1)},\, \nicefrac{3n - \lambda_1}{4n}^{(1)},\, \nicefrac{3n - \lambda_2}{4n}^{(1)}     \right\}$.
\qed

\subsection{Spectrum of the chopped correlation matrix $\Pi(2,2)$}

Lemma \ref{lem:pjell-easy} and Proposition \ref{lem1} provide the spectrum for all 
the chopped correlation matrices expect for 
$\Pi(2,2)$. We see that this chopped correlation matrix is quite complicated and 
a direct computation seems hopeless. We shall use the Heun operator introduced in Section \ref{sec:heun} commuting 
with $\Pi(2,2)$: we find the eigenvectors of the Heun operator and let $\Pi(2,2)$ act on them to get the eigenvalues of $\Pi(2,2)$.

As proved previously, the Heun operator \eqref{eq:T1} with $\nu=-Q_{1,1}-Q_{2,1} = -\sqrt{n}$ and $\mu=-P_{1,1}-P_{2,1} = -\sqrt{n}$ commutes with $\Pi(2,2)$ 
and reads as follows:
\begin{eqnarray}
  T&=& n^{\frac{3}{2}}E_0^* + n E_1^* - n E_3^* - n^{\frac{3}{2}} E_4^*  + (n+2\sqrt{n})\left(E_{0}^* A E_1^* + E_{1}^* A E_{0}^*\right)  \nonumber \\
 && +(2\sqrt{n}) \left(E_{1}^* A E_2^* + E_{2}^* A E_{1}^*\right) -n \left(E_{3}^* A E_4^* + E_{4}^* A E_{3}^*\right)\nonumber \\
 &=& \left( \begin{array}{c|c|c|c|c}	
            n^{\frac{3}{2}} & (n+2\sqrt{n}) \ones_n^T & \zeros & \zeros & \zeros \\
      \hline
			(n+2\sqrt{n}) \ones_n & nI_n & (2\sqrt{n}) M_1 & \zeros & \zeros \\
      \hline
			\zeros & (2\sqrt{n}) M_1^T & \zeros & \zeros & \zeros \\
      \hline
			\zeros & \zeros & \zeros & -nI_n & -n\ones_n \\
      \hline
			\zeros & \zeros & \zeros & -n \ones_n^T & -n^{\frac{3}{2}} \end{array} \right) \label{eq:T22}
\end{eqnarray}
where $M_1$ and $M_2$ are as in \eqref{eq:HadA1A3}. 
Since $T$ and $\Pi(2,2)$ commute, there exists a basis of $\re^{4n}$ which is a common eigenbasis for both matrices. Due to the block structure of both matrices, we need only diagonalise the first principal $3n-1 \times 3n-1$ submatrix of $T$.

\begin{proposition}\label{lem2} The spectrum of $\Pi(2,2)$ is
$\left\{ 0^{(n+1)},\frac{1}{4}^{(n-1)},1^{(2n-2)},\theta_1^{(1)}, \theta_2^{(1)} \right\}$ where
\[\theta_1 = \frac{3  n - \sqrt{5  n^{2} + 8  n^{\frac{3}{2}} - 4 \sqrt{n} - 5} - 1}{8 n} \text{ and } \theta_2 = \frac{3 n + \sqrt{5n^{2} + 8 n^{\frac{3}{2}} - 4 \sqrt{n} - 5} - 1}{8 n}. \]\end{proposition}

\proof Since $\Pi(2,2)$ has a $3n-1\times 3n-1$ block surrounded by zeroes, it has eigenvalues $0$ with multiplicity (at least) $n+1$. We will find the eigenvectors for the first  $3n-1\times 3n-1$ principal submatrix of $\Pi(2,2)$ by find an eigenbasis for the same principal submatrix of $T$ given in \eqref{eq:T22}. Since $T$ and $\Pi(2,2)$ commute by Corollary \ref{cor:Tcommutes}, they share an eigenbasis. In general, not every eigenbasis of $T$ need be an eigenbasis of $\Pi(2,2)$. we will find one that is. 

We will diagonalize the following matrix:
\[
T' = \left( \begin{array}{c|c|c}	
            n^{\frac{3}{2}} & (n+2\sqrt{n}) \ones_n^T & \zeros \\
      \hline
			(n+2\sqrt{n}) \ones_n & nI_n & (2\sqrt{n}) M_1  \\
      \hline
			\zeros & (2\sqrt{n}) M_1^T & \zeros \\
        \end{array} \right).
\]
Let 
\[
S =  \pmat{1& 0 &0 \\ 0 & \ones_n & 0 \\0& 0 &\ones_{2n-2}}.
\]
Since each block of $T'$ has constant rows, we will find eigenvalues of the matrix:
\[
B := \pmat{1& 0 &0 \\ 0 & \frac{1}{n} & 0 \\0& 0 &\frac{1}{2n-2}} S^T T' S =  \pmat{n^{\nicefrac{3}{2}} & n^2 + 2 n^{\nicefrac{3}{2}} &0 \\ n + 2 n^{\nicefrac{1}{2}} & n & 2n^{\nicefrac{1}{2}}(n-1) \\0& n^{\nicefrac{3}{2}} &0}.
\]
Consider an eigenvector $\Zv$ of $B$ with eigenvalue $\lambda$. It is easy to see that the first entry of $\Zv$ is non-zero (else $\Zv = \zeros$) so we may scale so that $\Zv = \pmat{1&a&b}^T$ for some $a,b$. We obtain that
\[ B\Zv = \pmat{n^{\nicefrac{3}{2}} + a\left(n^2 + 2 n^{\nicefrac{3}{2}}\right) \\ n + 2 n^{\nicefrac{1}{2}} + an + 2bn^{\nicefrac{1}{2}}(n-1) \\ an^{\nicefrac{3}{2}}}  = \lambda \Zv .
\]
Thus we have that
\begin{equation}\label{eq:Beigs-ab}  a(
\lambda) = \frac{\lambda - n^{\nicefrac{3}{2}}}{n^2 + 2n^{\nicefrac{3}{2}}}, \quad b(\lambda) = \frac{an^{\nicefrac{3}{2}}}{\lambda} .
\end{equation}
We need only find the eigenvalues of $B$ and obtain their eigenvector using \eqref{eq:Beigs-ab}. Observe that 
\[\lambda_1 = 2n^{\nicefrac{3}{2}} + 2n, \quad a(\lambda_1) = n^{-\nicefrac{1}{2}}, \quad b(\lambda_1) = \frac{1}{2n^{\nicefrac{1}{2}}+2} \] 
is one such solution. Let $\mu, \nu$ be the two other eigenvalues of $B$. We have that 
\[
\mu\nu = \frac{\det B}{\lambda_1} = -n^3 + n^{\nicefrac{5}{2}} \text{ and } \mu + \nu = \tr B - \lambda_1 = - n^{\nicefrac{3}{2}} - n. 
\]
Thus, $\mu,\nu$ are the two solution of $
(x-\mu)(x-\nu) = x^2 - (\mu+\nu)x + \mu\nu$, and thus 
\begin{equation}
\mu = \frac{-\left(n^{\nicefrac{3}{2}}+ n\right) + \sqrt{5n^3 - 2n^{\nicefrac{3}{2}} + n^2}}{2},\,
\nu = \frac{-\left(n^{\nicefrac{3}{2}}+ n\right) - \sqrt{5n^3 - 2n^{\nicefrac{3}{2}} + n^2}}{2}.
\end{equation} 
We have thus the following three eigenvectors for $T'$
\begin{equation}\label{eq:Teigslmn}
\Zu_0 = \pmat{1 \\n^{-\nicefrac{1}{2}}\ones_n \\  \frac{1}{2n^{\nicefrac{1}{2}}+2}\ones_{2n-2} } , \quad
\Zu_1 =\pmat{1 \\ a(\mu) \ones_n \\   b(\mu) \ones_{2n-2} } , \quad
\Zu_2 =\pmat{1 \\ a(\nu)\ones_n \\ b(\nu) \ones_{2n-2} } 
\end{equation}
with eigenvalues $\lambda_1$, $\mu$ and $\nu$, respectively.

Let $\Zf_0,\ldots, \Zf_{n-2}$ be the elementary basis for $\re^{n-1}$; we mean that $\Zf_i$ is the $(n-1)$-dimensional vector with $1$ in the $i$th position and $0$ elsewhere. We observe that 
\[
M_1 \left(\pmat{1 \\1} \otimes \pmat{\Zf_i - \Zf_{n-2}}\right) = 0
\]
for $i = 0,\ldots, n-3$. Thus, we have that
\[
\Zv_i = \pmat{ 0 \\ \zeros_{n} \\\pmat{1 \\1} \otimes \pmat{\Zf_i - \Zf_{n-2}}} 
\]
is an eigenvector of $T'$ with eigenvector $0$ for $i = 0,\ldots, n-3$. It is also easy to observe that $\{\Zv_i\}_{i=0}^{n-3}$ is a linearly independent set since $\{f_i - f_{n-2}\}_{i=0}^{n-3}$ is linearly independent. 

Let $\Ze_0,\ldots, \Ze_{n-2}$ be the elementary basis for $\re^{n-1}$
Consider the following:
\[
T' \pmat{0 \\ \Ze_i - \Ze_{n-1} \\ \Zx } =   \pmat{0 \\ n(\Ze_i - \Ze_{n-1}) + 2\sqrt{n}M_1 \Zx \\ 2\sqrt{n}M_1^T (\Ze_i - \Ze_{n-1}) }. 
\]
Observe that 
\[
 2\sqrt{n} 2\sqrt{n} M_1 M_1^T (\Ze_i - \Ze_{n-1}) = 2 n (nI_n - (n-2) J_n)  (\Ze_i - \Ze_{n-1}) = 2n^2  (\Ze_i - \Ze_{n-1}). 
\]
For this, we can find vectors $\Zx$ which give eigenvectos of $T'$. Let 
\[
\Zw_i = \pmat{0 \\ \Ze_i - \Ze_{n-1} \\ -2n^{-\nicefrac{1}{2}} M_1^T (\Ze_i - \Ze_{n-1}) } \text{ and } \Zz_i = \pmat{0 \\ \Ze_i - \Ze_{n-1} \\ n^{-\nicefrac{1}{2}} M_1^T (\Ze_i - \Ze_{n-1}) }
\]
for $i =0, \ldots n-2$. We see that $T'\Zw_i = -n \Zw_i$ and $T'\Zz_i = 2n \Zz_i$. Since $\{\Ze_i - \Ze_{n-1} \}_{i=0}^{n-2} $ is a linearly independent set, as are $\{\Zw_i\}_{i=0}^{n-2} $ and $\{\Zz_i\}_{i=0}^{n-2} $. 

In summary, the eigenvectors of $T'$ are $\Zu_0,\Zu_1,\Zu_2$,   $\{\Zv_i\}_{i=0}^{n-3}$, $\{\Zw_i\}_{i=0}^{n-2} $ and $\{\Zz_i\}_{i=0}^{n-2} $. Now we will find the eigenvalues of $\Pi(2,2)$. Let $P'$ be the first $3n-1 \times 3n-1$ principal submatrix of $\Pi(2,2)$. 

We can compute that $P' \Zu_0 = \Zu_0$ and the eigenvalues of $P'$ corresponding to $\Zu_1$ and $\Zu_2$ are 
\[\theta_1 = \frac{3 \, n - \sqrt{5 \, n^{2} + 8 \, n^{\frac{3}{2}} - 4 \, \sqrt{n} - 5} - 1}{8 \, n} \text{ and } \theta_2 = \frac{3 \, n + \sqrt{5 \, n^{2} + 8 \, n^{\frac{3}{2}} - 4 \, \sqrt{n} - 5} - 1}{8 \, n}. \]

Since $f_i - f_{n-2}$ is orthogonal to each row of $J_{n-1}$ and 
\[  \left( \begin{array}{c}  J_{n-1,n}+ \sqrt{n}\comp{H}^T \\ \hline  J_{n-1,n} - \sqrt{n}\comp{H}^T\end{array} \right)\]
we obtain that $P'\Zv_i = \Zv_i$ for all $i = 0,\ldots, n-3$. 

Similarly, we can find that $P'\Zw_i = \frac{1}{4} \Zw_i$ and $P'\Zz_i = \Zz_i$. \qed

\subsection{Entanglement entropy for Hadamard graph}

We have just computed analytically the eigenvalues $\{\nu_j\}$ of the chopped correlation matrix $C= \Pi(K,\ell)$. From them we can compute the von Neumann entropy by using the relation
\begin{equation}
    S_{K,\ell}= -\sum_{j} \Big(\nu_j \ln(\nu_j)+(1-\nu_j)\ln(1-\nu_j) \Big)\,.
\end{equation}
From the results of Propositions \ref{lem1} and \ref{lem2}, we can determine exactly the entanglement entropy for different configurations. In particular, the following exact asymptotics are found as a function of the size $n$ of the Hadamard matrix:
\begin{eqnarray}
&& S_{2,3}\sim S_{1,3}  \underset{n\rightarrow \infty}{\sim}  2\ln(2)-\frac{3}{4}\ln(3)_,,\label{1}\\
&& S_{3,3}  \underset{n\rightarrow \infty}{\sim} \frac{\ln(n) }{4n}_,,\label{2}\\
&&S_{1,1}  \sim S_{1,2}  \sim S_{2,2}  \underset{n\rightarrow \infty}{\sim} 
\left(2\ln(2)-\frac{3}{4}\ln(3)\right)n\,.\label{3}
\end{eqnarray}

\begin{figure}
    \centering
    \includegraphics{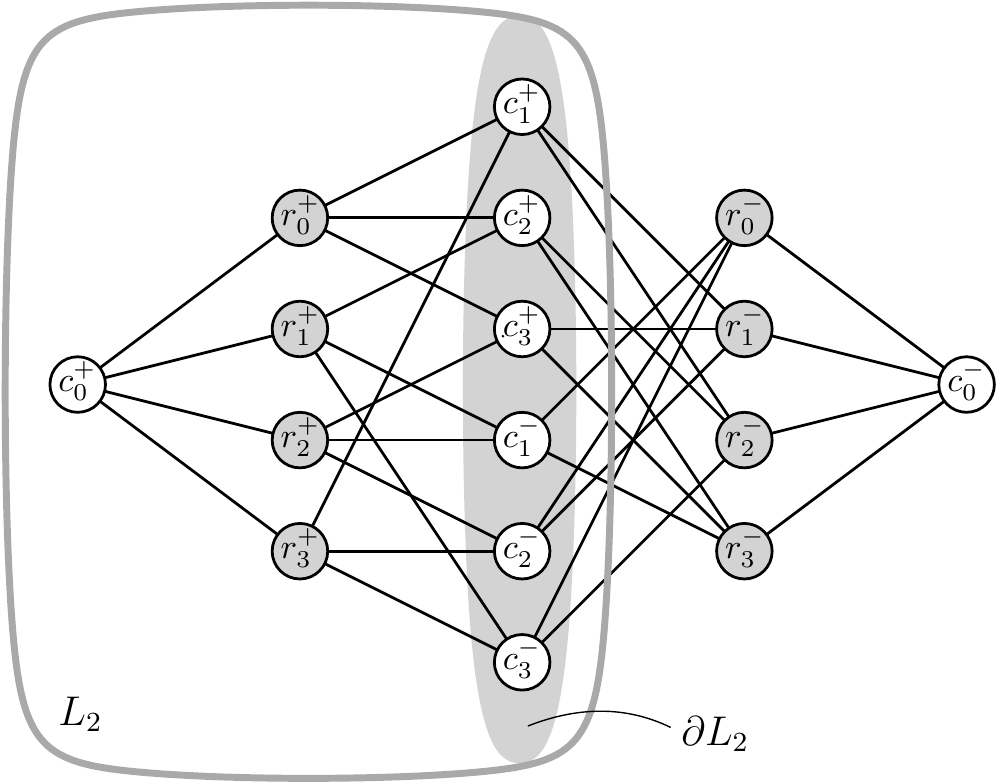}
    \caption{The Hadamard graph, $\cH_4$, of the $4\times 4$ Hadamard matrix $H_4$ with $L_2$ and $\partial L_2$ displayed. \label{fig:had4-L2}}
\end{figure}

These results can be qualitatively understood as follows. We should keep in mind that the adjacency matrix $A_1$ as a Hamiltonian is non-local. Now the Hadamard graph $\mathcal{H}_n$ has $4n$ vertices. While a given site does not interact with all sites it is nevertheless coupled to $n$ of them, that is to a number equal to the order of the graph. Let $L_\ell=Im (\Pi_1(\ell))$ be one part of the bipartition and define its volume by $\mathcal{V}(L_\ell )=Card (L_\ell) \; (=N_1)$. The boundary of $L_\ell \;is \;\partial L_\ell=\{v_i \in L_\ell : \exists\; v_j \in \mathcal{H}_n\smallsetminus L_\ell \;\text{with} \; d(v_1,v_j)=1\}$. Figure \ref{fig:had4-L2} shows $L_2$ and its boundary on the Hadamard graph $\cH_4$. The area $\mathcal{S}(L_\ell)$ would be defined as $\mathcal{S}(L_\ell) = Card (\partial L_\ell)$, see \cite{Eis10}. If $l=1, 2, 3$, from the definition of the Hadamard graph, it is immediate to see that $\mathcal{V}(L_\ell )$ and $\mathcal{S}(L_\ell )$ are linear functions of $n$. The distinction between area and volume law in the thermodynamic limit is therefore pointless in this case.

Consider formula \eqref{1}. The part $L_3$ of the graph contains all but one of its vertices. It is hence not a surprise that the distribution of quantum correlations does not depend on $n$ in this situation. Looking at \eqref{2}, the Fermi sea in this case extends to almost all modes. The correlation matrix $C=\Pi_1(\ell) \Pi_2(K) \Pi_1(\ell)$ does then get closer and closer to $1$ (like $\Pi_2(K))$ as $n$ becomes large. We thus realize that its eigenvalues will all approach $1$ and that the entropy will go to zero as $n$ grows. Finally, for $\ell =1, 2$, in view of the non-local character of the Hamiltonian, it is expected that the entropy will increase like the volume equivalently, like $n$, in the thermodynamic limit as is confirmed by \eqref{3}.

\section{Concluding Remarks}

This paper has explored the entanglement of free Fermions on graphs of association schemes. It has shown how a block-tridiagonal operator that commutes with the entanglement Hamiltonian can be determined in certain instances. This relied on the extension of the algebraic Heun construct to the realm of Terwilliger algebras. The formalism has been applied to the study of Fermions on Hadamard graphs. The existence of the commuting operator has proved its use in this case as it allowed to diagonalize exactly the associated entanglement Hamiltonian - an achievement which is not so common.

Looking forward, the examination of the entanglement of free Fermions (or harmonic oscillators) on graphs of other associations schemes such as the Hamming and Johnson ones obviously comes to mind. With an eye to wider applications of algebraic Heun operators, we feel it would be most interesting to explore how such constructions could be further extended to tridiagonal pairs and algebras and how this could then also apply to entanglement studies. We plan on pursuing some of these questions.

\subsection*{Acknowledgments}

The authors are grateful to William Witczak-Krempa for very useful exchanges. The research of LV is funded in part by a
Discovery Grant from the Natural Sciences and Engineering Research Council (NSERC) of Canada. 

\bibliographystyle{plain}

\begin{thebibliography}{1}

\bibitem{BCN}
A.~E. Brouwer, A.~M. Cohen and A.~Neumaier,
\textsl{Distance-regular graphs},
Springer-Verlag, Berlin, 1989.

\bibitem{CFG} J.~A. Carrasco, F. Finkel, A. Gonz\`alez-L\'opez and P. Tempesta, 
\textsl{A duality principle for the multi-block entanglement entropy of free fermion systems,} 
Scientific Reports 7 (2017) 11206 and \texttt{arXiv:1701.05355}.

\bibitem{CoutinhoGodsilGuoVanhove2}
G.~Coutinho, C.~Godsil, K.~Guo and F.~Vanhove,
\textsl{Perfect state transfer on distance-regular graphs and association
  schemes,}
Linear Algebra and its Applications 478 (2015) 108 -- 130 and \texttt{arXiv:1401.1745}.

\bibitem{CNV1}N. Crampe, R.~I. Nepomechie and L.Vinet,
\textsl{Free-Fermion entanglement and orthogonal polynomials,}
J. Stat. Mech. (2019) 093101 and \texttt{arXiv:1907.00044}.

\bibitem{CNV2} N. Crampe, R.~I. Nepomechie and L.Vinet,
\textsl{Entanglement in Fermionic Chains and Bispectrality,}
Roman Jackiw 80th Birthday Festschrift, A. Niemi, T. Tomboulis, K. K. Phua eds.
World Scientific, pp. 77-96, (2020) and
\texttt{arXiv:2001.10576}.

\bibitem{drgsurvey}
E. van Dam, J. Koolen and H. Tanaka,
\textsl{Distance-regular graphs,}
The electronic journal of combinatorics 1 (2014) 10 and \texttt{arXiv:1410.6294v2}.

\bibitem{DG}
J.~J. Duistermaat and F.~A. Gr\"unbaum,
\textsl{Differential equations in the spectral parameter,}
Commun. Math. Phys. 103 (1986) 177--240.

\bibitem{Eis10}
J.~{Eisert}, M.~{Cramer} and M.~B. {Plenio},
\textsl{Area laws for the entanglement entropy - a review}, 
Rev. Mod. Phys. 82 (2010) 277 and \texttt{arXiv:0808.3773}.

\bibitem{EP04}
V. Eisler and I. Peschel,
\textsl{Free-fermion entanglement and spheroidal functions,}
J. Stat. Mech. (2013) P04028 and \texttt{arXiv:1302.2239 }.

\bibitem{Eisler18}
V. Eisler and I. Peschel,
\textsl{
Properties of the entanglement Hamiltonian for finite free-fermion chains,}
J. Stat. Mech. (2018) 104001 and 
\texttt{arXiv:1805.00078}.

\bibitem{GK}
D. Gioev and I. Klich, 
\textsl{Entanglement entropy of fermions in any dimension and the Widom conjecture,}
Phys. Rev. Lett. 96 (2006) 100503 and \texttt{arXiv:quant-ph/0504151}.

\bibitem{GO} J.~T. Go,
\textsl{The Terwilliger Algebra of the Hypercube,}
Europ. J. Combinatorics  23 (2002) 399--429.

 \bibitem{G}
 F.~A. Gr\"unbaum,  
 \textsl{Time‐band limiting and the bispectral problem,}
 Communications on Pure and Applied Mathematics 47 (1994) 307--328.
 
 \bibitem{GPZ}
 F.~A. Gr\"unbaum, I. Pacharoni and I. Zurrián,
 \textsl{Bispectrality and Time–Band Limiting: Matrix-valued Polynomials,}
 International Mathematics Research Notices 2020 (2020) 4016--4036 and \texttt{arXiv:1801.10261}.


\bibitem{GVZ1}
 F.~A. Gr\"unbaum, L. Vinet and A. Zhedanov, 
 \textsl{Algebraic Heun Operator and Band-Time Limiting,}
 Commun. Math. Phys. 364 (2018) 1041--1068 and \texttt{arXiv:1711.07862}
 
\bibitem{GVZ2}
 F.~A. Gr\"unbaum, L. Vinet and A. Zhedanov, 
 \textsl{Tridiagonalization and the Heun equation,}
 J.  Math. Phys. 58 (2017) 031703 and \texttt{arXiv:1602.04840}.
 
\bibitem{2012PhRvB..86x5109H}
Z.~{Huang} and D.~P. {Arovas}, 
\textsl{Entanglement spectrum and Wannier center flow   of the Hofstadter problem}, 
  Phys. Rev. B 86 (2012) 245109 and \texttt{arXiv:1201.0733}.
  
\bibitem{IKT} 
T. Ito, K. Tanabe and P. Terwilliger, \textsl{Some algebra related to P- and Q-polynomial association schemes,} 
in: Codes and Association Schemes (Piscataway NJ, 1999), Amer. Math. Soc., Providence RI, 2001, pp. 167--192 and \texttt{arXiv:math.CO/0406556}.

\bibitem{JES1}
M.~A. Jafarizadeh, F. Eghbalifam and S. Nami,
\textsl{Entanglement entropy of free fermions on directed graphs,}
Eur. Phys. J. Plus 132 (2017) 1--13.

\bibitem{JES2}
M.~A. Jafarizadeh, F. Eghbalifam and S. Nami,
\textsl{Entanglement entropy in the spinless free fermion model and its application to the graph isomorphism problem,}
J. Phys. A 51 (2018) 075304.

\bibitem{Landau}
H.~J. Landau,
\textsl{An overview of time and frequency limiting,} in Fourier techniques and applications Springer, (1985) pp. 201--220.

\bibitem{LR}
J.~I. Latorre and A. Riera,
\textsl{
A short review on entanglement in quantum spin systems,}
J. Phys A 42 (2009) 504002 and \texttt{arXiv:0906.1499}.

\bibitem{Lee:2014nra}
C.~H. Lee, P.~Ye and X.-L. Qi, 
\textsl{Position-momentum duality in the entanglement spectrum of free fermions}, 
  J. Stat. Mech. 1410 (2014) P10023 and \texttt{arXiv:1403.1039}.
  
 \bibitem{NT}
 K. Nomura and P. Terwilliger,
 \textsl{Linear transformations that are tridiagonal with respect to both eigenbases of a Leonard pair,}
 Linear algebra and its applications, 420(1), (2007) 198--207 and \texttt{arXiv:math/0605316}.
 
 \bibitem{RKP}
 R.~K. Perline, 
 \textsl{ Discrete time-band limiting operators and commuting tridiagonal matrices,} SIAM Journal on Algebraic Discrete Methods 8 (1987) 192--195.
  
  \bibitem{2003JPhA...36L.205P} I.~{Peschel}, 
\textsl{Calculation of reduced density matrices
  from correlation functions}, 
  J. Phys A 36 (2003) L205--L208 and \texttt{arXiv:cond-mat/0212631}.
  
  \bibitem{Peschel04}
  I. Peschel, 
 \textsl{ On the reduced density matrix for a chain of free electrons,}
  J. Stat. Mech. (2004) P06004 and \texttt{arXiv:cond-mat/0403048}.
  
  \bibitem{Peschel12}
  I. Peschel, 
  \textsl{
  Special review: Entanglement in solvable many-particle models,}
 	Braz. J. Phys. 42 (2012) 267--291 and \texttt{arXiv:1109.0159}.


\bibitem{2009JPhA...42X4003P}
I.~{Peschel} and V.~{Eisler}, 
\textsl{Reduced density matrices and entanglement
  entropy in free lattice models}, 
  J. Phys. A 42 (2009) 504003 and \texttt{arXiv:0906.1663}.
  
  
\bibitem{P94} V.~V. Prasolov,
\newblock {\textsl{ Problems and theorems in linear algebra}}, volume 134 of {
  Translations of Mathematical Monographs}.
\newblock American Mathematical Society, Providence, RI, 1994.
\newblock Translated from the Russian manuscript by D. A. Le{\u\i}tes.
  
 \bibitem{Slepian83}
 D. Slepian, 
 \textsl{Some comments on Fourier analysis, uncertainty and modeling,}
 SIAM review 25 (1983) 379--393.

  \bibitem{SP61}
 D. Slepian and H.~D. Pollak,  
 \textsl{Prolate spheroidal wave functions, Fourier analysis and uncertainty—I,}
 Bell System Technical Journal 40 (1961) 43--63.
  
 \bibitem{Ter92}
P. Terwilliger,
\textsl{The subconstituent algebra of an association scheme, (part i),}
Journal of Algebraic Combinatorics 1 (1992) 363--388.

\bibitem{T03}
P. Terwilliger, 
\textsl{Introduction to Leonard pairs,}
Journal of Computational and Applied mathematics 153 (2003) 463--475.


\end{thebibliography}

\end{document}